\documentclass[sigconf]{acmart}

\makeatletter
\def\@ACM@checkaffil{% Only warnings
    \if@ACM@instpresent\else
    \ClassWarningNoLine{\@classname}{No institution present for an affiliation}%
    \fi
    \if@ACM@citypresent\else
    \ClassWarningNoLine{\@classname}{No city present for an affiliation}%
    \fi
    \if@ACM@countrypresent\else
        \ClassWarningNoLine{\@classname}{No country present for an affiliation}%
    \fi
}
\makeatother
\AtBeginDocument{%
}

\copyrightyear{2025}
\acmYear{2025}
\setcopyright{rightsretained}
\acmConference[FPGA '25]{Proceedings of the 2025 ACM/SIGDA International Symposium on Field Programmable Gate Arrays}{February 27--March 1, 2025}{Monterey, CA, USA}
\acmBooktitle{Proceedings of the 2025 ACM/SIGDA International Symposium on Field Programmable Gate Arrays (FPGA '25), February 27--March 1, 2025, Monterey, CA, USA}
\acmDOI{10.1145/3706628.3708873}
\acmISBN{979-8-4007-1396-5/25/02}


\usepackage{balance}

\usepackage{tikz}
\usetikzlibrary{shapes,arrows,positioning}
\usepackage{listings} % For code listings
\usepackage{caption}  % For captioning the listings
\usepackage{graphicx} % For flexible width setting
\usepackage{xcolor}
\usepackage{dsfont}
\usepackage{geometry}
\usepackage{changepage}
\usepackage{tikz}
\usepackage{pifont}
\usepackage{caption}
\usetikzlibrary{tikzmark}
\usepackage{tabularx}
\newcommand{\myparagraph}[1]{\par\textbf{\textit{ #1}}~}
\newcommand{\myparagraphee}[1]{\par\textit{ #1}~}
\usepackage{multicol}
\usepackage{enumitem}
\usepackage{longtable}
\usepackage{supertabular}
\definecolor{codegreen}{rgb}{0,0.6,0}
\definecolor{codegray}{rgb}{0.5,0.5,0.5}
\definecolor{codepurple}{rgb}{0.58,0,0.82}
\definecolor{backcolour}{rgb}{0.95,0.95,0.92}

\usepackage{setspace}  % Include this in the preamble

\lstdefinestyle{cppstyle}{
    language=C++,
    commentstyle=\color{codegreen},
    keywordstyle=\color{blue},
    numberstyle=\tiny\color{codegray},
    stringstyle=\color{codepurple},
    basicstyle=\fontsize{6}{8}\ttfamily\scriptsize, 
    breakatwhitespace=false,
    breaklines=true,
    captionpos=b,
    keepspaces=true,
    numbers=left,
    numbersep=5pt,
    showspaces=false,
    showstringspaces=false,
    showtabs=false,
    tabsize=2,
    aboveskip=1pt, % Espace au-dessus des listings
    belowskip=1pt, % Espace en dessous des listings
    xleftmargin=0.5cm
}

\usepackage[caption=false, font=normalsize, labelfont=sf, textfont=sf]{subfig}

\usepackage{xspace}
\newcommand{\framework}{Sisyphus\xspace}

\newcommand{\yes}{\textcolor{green}{\ding{51}}}
\newcommand{\no}{\textcolor{red}{\ding{55}}}

\newcommand{\yesbad}{\textcolor{red}{\ding{51}}}
\newcommand{\nogood}{\textcolor{green}{\ding{55}}}

\let\oldtable\table
\let\endoldtable\endtable
\renewenvironment{table}[1][ht]{%
  \oldtable[#1]\small
}{%
  \endoldtable
}
\makeatletter
\g@addto@macro\normalsize{%
    \setlength{\abovedisplayskip}{1pt}
    \setlength{\belowdisplayskip}{1pt}
    \setlength{\abovedisplayshortskip}{1pt}
    \setlength{\belowdisplayshortskip}{1pt}
}
\makeatother
\usepackage{caption}
\makeatletter
\gdef\@copyrightpermission{
 \begin{minipage}{0.2\columnwidth}
  \href{https://creativecommons.org/licenses/by/4.0/}{\includegraphics[width=0.90\textwidth]{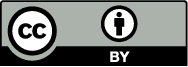}}
 \end{minipage}\hfill
 \begin{minipage}{0.8\columnwidth}
  \href{https://creativecommons.org/licenses/by/4.0/}{This work is licensed under a Creative Commons Attribution International 4.0 License.}
 \end{minipage}
 \vspace{5pt}
}
\makeatother

\begin{document}

%%
%% The "title" command has an optional parameter,
%% allowing the author to define a "short title" to be used in page headers.
\title{A Unified Framework for Automated Code Transformation and Pragma Insertion}

%%
%% The "author" command and its associated commands are used to define
%% the authors and their affiliations.
%% Of note is the shared affiliation of the first two authors, and the
%% "authornote" and "authornotemark" commands
%% used to denote shared contribution to the research.
\author{Stéphane Pouget}
\email{pouget@cs.ucla.edu}
\orcid{0000-0003-3950-5818}
% \author{G.K.M. Tobin}
% \authornotemark[1]
% \email{webmaster@marysville-ohio.com}
\affiliation{%
\institution{University of California, Los Angeles}
% \department{Henry Samueli School of Engineering}
% \streetaddress{Samuels Building (F25), Kensington Campus}
\city{Los Angeles}
\state{CA}
% \postcode{2052}
\country{USA}}

\author{Louis-Noël Pouchet}
\email{pouchet@colostate.edu}
\orcid{0000-0001-5103-3097}
\affiliation{%
\institution{Colorado State University}
% \department{Department of Computer Science}
% \streetaddress{Samuels Building (F25), Kensington Campus}
\city{Fort Collins}
\state{CO}
% \postcode{2052}
\country{USA}}

\author{Jason Cong}
\email{cong@cs.ucla.edu}
\orcid{0000-0003-2887-6963}
\affiliation{%
\institution{University of California, Los Angeles}
% \department{Henry Samueli School of Engineering}
% \streetaddress{Samuels Building (F25), Kensington Campus}
\city{Los Angeles}
\state{CA}
% \postcode{2052}
\country{USA}}

%%
%% By default, the full list of authors will be used in the page
%% headers. Often, this list is too long, and will overlap
%% other information printed in the page headers. This command allows
%% the author to define a more concise list
%% of authors' names for this purpose.
% \renewcommand{\shortauthors}{Pouget et al.}
\renewcommand{\shortauthors}{Stéphane Pouget, Louis-Noël Pouchet, Jason Cong}

%%
%% The abstract is a short summary of the work to be presented in the
%% article.
\begin{abstract}
High-Level Synthesis compilers and Design Space Exploration tools have greatly advanced the automation of hardware design, improving development time and performance. However, achieving a good Quality of Results still requires extensive manual code transformations, pragma insertion, and tile size selection, which are typically handled separately. The design space is too large to be fully explored by this fragmented approach. It is too difficult to navigate this way, limits the exploration of potential optimizations, and complicates the design generation process.

To tackle this obstacle, we propose Sisyphus, a unified framework that automates code transformation, pragma insertion, and tile size selection within a common optimization framework. By leveraging Nonlinear Programming, our approach efficiently explores the vast design space of regular loop-based kernels, automatically selecting loop transformations and pragmas that minimize latency. 

Evaluation against state-of-the-art frameworks, including AutoDSE, NLP-DSE, and ScaleHLS, shows that Sisyphus achieves superior Quality of Results, outperforming alternatives across multiple benchmarks. By integrating code transformation and pragma insertion into a unified model, Sisyphus significantly reduces design generation complexity and improves performance for FPGA-based systems.

% High-level synthesis, source-to-source compilers, and various Design Space Exploration techniques for pragma insertion have significantly improved the Quality of Results of generated designs. These tools offer benefits such as reduced development time and enhanced performance. However, achieving high-quality results often requires additional manual code transformations and tiling selections, which are typically performed separately or as pre-processing steps. Although DSE techniques enable code transformation upfront, the vastness of the search space often limits the exploration of all possible code transformations, making it challenging to determine which transformations are necessary. Additionally, ensuring correctness remains challenging, especially for complex transformations and optimizations.

% To tackle this obstacle, we first propose a comprehensive framework leveraging HLS compilers. Our system streamlines code transformation, pragma insertion, and tiles size selection for on-chip data caching through a unified optimization problem, aiming to enhance parallelization, particularly beneficial for computation-bound kernels. Them employing a novel Nonlinear Programming (NLP) approach, we simultaneously ascertain transformations, pragmas, and tile sizes, focusing on regular loop-based kernels. Our evaluation demonstrates that our framework adeptly identifies the appropriate transformations, including scenarios where no transformation is necessary, and inserts pragmas to achieve a favorable Quality of Results.

% 

\end{abstract}

%%
%% The code below is generated by the tool at http://dl.acm.org/ccs.cfm.
%% Please copy and paste the code instead of the example below.
%%
% \begin{CCSXML}
% <ccs2012>
%  <concept>
%   <concept_id>00000000.0000000.0000000</concept_id>
%   <concept_desc>Do Not Use This Code, Generate the Correct Terms for Your Paper</concept_desc>
%   <concept_significance>500</concept_significance>
%  </concept>
%  <concept>
%   <concept_id>00000000.00000000.00000000</concept_id>
%   <concept_desc>Do Not Use This Code, Generate the Correct Terms for Your Paper</concept_desc>
%   <concept_significance>300</concept_significance>
%  </concept>
%  <concept>
%   <concept_id>00000000.00000000.00000000</concept_id>
%   <concept_desc>Do Not Use This Code, Generate the Correct Terms for Your Paper</concept_desc>
%   <concept_significance>100</concept_significance>
%  </concept>
%  <concept>
%   <concept_id>00000000.00000000.00000000</concept_id>
%   <concept_desc>Do Not Use This Code, Generate the Correct Terms for Your Paper</concept_desc>
%   <concept_significance>100</concept_significance>
%  </concept>
% </ccs2012>
% \end{CCSXML}

% \ccsdesc[500]{Do Not Use This Code~Generate the Correct Terms for Your Paper}
% \ccsdesc[300]{Do Not Use This Code~Generate the Correct Terms for Your Paper}
% \ccsdesc{Do Not Use This Code~Generate the Correct Terms for Your Paper}
% \ccsdesc[100]{Do Not Use This Code~Generate the Correct Terms for Your Paper}

%%
%% Keywords. The author(s) should pick words that accurately describe
%% the work being presented. Separate the keywords with commas.

%% A "teaser" image appears between the author and affiliation
%% information and the body of the document, and typically spans the
%% page.

% \received{20 February 2007}
% \received[revised]{12 March 2009}
% \received[accepted]{5 June 2009}

%%
%% This command processes the author and affiliation and title
%% information and builds the first part of the formatted document.
\maketitle

\section{Introduction}
\label{introduction}

High-Level Synthesis (HLS) compilers and source-to-source compilers are indispensable tools in accelerating hardware design by automating the translation of high-level programming languages like C/C++ or Python into hardware descriptions. They offer notable advantages, such as reduced development time and enhanced performance for hardware designs \cite{cong.tcadics.2011, Zhang2008, vitis, intel_fpga, catapult, legup, merlin, pylog, scalehls, heterocl, s2fa, heterohalide,pom,10.1145/3530775}. However, achieving a good quality of results (QoR) often requires manual code transformations and pragma insertions, which can be facilitated with either Design Space Exploration (DSE) \cite{nlp_dse_poster, nlp_dse, autodse, harp, sohrabizadeh2022gnn, lorenzo, comba, harp} or source-to-source compilers \cite{heterocl, scalehls,hida,pom} to guide the synthesis process. These pragmas aid in optimizing the generated hardware code. 

Despite their advantages, existing HLS workflows face several significant challenges. Previous frameworks \cite{scalehls, hida, heterocl,pom} have integrated pragma insertion and code transformation, but these efforts are somewhat limited, focusing primarily on loop permutations based on loop properties. Additionally, code transformations and pragma insertion are often treated as separate processes. Transformations such as loop tiling, splitting, and permutation are typically performed in advance or as part of a pre-processing stage, followed by pragma insertion for optimization, like loop unrolling, pipelining and array partitioning.
% while pragmas for optimizations like loop unrolling, pipelining, and array partitioning are applied later. 
% 
This separation introduces inefficiencies, as transformations can affect the effectiveness of pragmas, and vice versa.
Moreover, the immense design space—encompassing millions of potential transformation and pragma combinations—makes exhaustive exploration unfeasible. Ensuring the correctness of these transformations while adhering to resource constraints further complicates the process.

Our primary research objective is to develop a system capable of autonomously conducting code transformation, tile size selection to cache the data on-chip and integrating hardware pragmas for HLS to enhance parallelization. 
We seek to attain a favorable QoR, especially for computation-bound designs, where maximizing parallelism is crucial for optimizing QoR.
To address this challenge, we introduce \framework, a framework built on top of HLS compilers. This framework automates code transformation, including loop splitting, permutation, tiling, and pragma insertion for unrolling, pipelining, and on-chip data caching while partitioning arrays to ensure efficient parallelization, all within a single optimization framework.
Even though several exploration methods could be used, we chose to employ a Nonlinear Programming (NLP) approach. This method facilitates rapid exploration of the entire theoretical space.
We have developed an analytical model that integrates considerations of latency and resource utilization, building upon previous research \cite{nlp_dse, nlp_dse_poster}. Our focus lies specifically on regular loop-based kernels \cite{polybench-web}, ensuring meticulous control over the correctness of code transformations and the accuracy of cost models. This model relies on parameters derived from both the program's schedule and the pragma configuration.
To facilitate seamless code transformation while respecting constraints and pragma insertion, we have designed a novel template tailored to these objectives.
In our template, we incorporate a two-level tiling strategy, where each tiling level corresponds to particular HLS optimizations such as fine-grained unrolling, pipelining, coarse-grained unrolling, or tiling. As a result, the loop trip counts become variables, with each loop associated with a specific pragma determined by the tiling level. By solving the NLP problem, we determine these trip counts, as well as other parameters like the array partitioning factor, enabling the automatic generation of the corresponding C++ code.

\noindent In summary, we introduce the following contributions:
% \vspace{-0.15cm}
\begin{itemize}[leftmargin=*]

\item A novel optimization template that integrates code transformation, pragma insertion, and tiles size selection for data caching by consolidating these tasks into a single optimization problem. This approach not only simplifies the search space but also ensures that only legal transformations are considered.

\item A novel NLP-based approach specifically designed to explore the joint design space of regular loop-based kernels. Unlike traditional methods that involve time-consuming DSE, our approach provides solutions within 
% seconds or 
minutes. Plus, the NLP is user-friendly, allowing users to easily configure parameters, such as on-chip buffer size, to meet the specific requirements of the kernel.

\item Our framework acts as a comprehensive, fully automated system, offering end-to-end functionality. With it, we can conduct thorough evaluations and attain QoR designs that are comparable or superior to those achieved by AutoDSE, NLP-DSE, ScaleHLS and HARP -- 
\textit{all without requiring DSE to perform multiple HLS compiler syntheses}.
% \textit{all without the necessity of DSE}. 
% Furthermore, \framework accurately identifies the appropriate transformation, even in cases where no transformation is required.
Furthermore, \framework identifies the correct transformation, 
even when no transformation is needed.
% even when none is needed.
\end{itemize}

% \vspace{-0.2cm}

The paper is organized into the following sections. Section~\ref{sec:motivation} 
% lays out the motivation behind our approach and the proposed solution.
explains our approach and solution. 
Following that, Section~\ref{sec:code_transformation} 
% presents the range of code transformation and pragma insertion we consider. 
covers code transformation and pragma insertion.
Section~\ref{sec:nlp} introduces a nonlinear formulation based on this model to automatically discover schedule and pragma configurations through NLP optimization.  
% Section~\ref{sec:nlp} introduces a nonlinear formulation to discover schedule and pragma configurations via NLP optimization.
Section~\ref{sec:code_generation} elaborates on our post-optimizations. Finally, sections \ref{sec:evaluation}, \ref{sec:related}, \ref{sec:limitation}  and \ref{sec:conclusion} are dedicated to evaluating our method, presenting related work, addressing limitations and drawing conclusions.

\section{Background and Motivation}
\label{sec:motivation}

\subsection{Design Space to Explore}

The exploration space for optimizing an HLS design includes pragma insertion, code transformations, and data caching, all within the limits of resource constraints. Although different objective functions can be explored, this work specifically focuses on minimizing latency while respecting resource constraints.
This objective function is especially potent for computation bound kernels because it addresses the bottleneck caused by insufficient parallelization.
% 
% These different elements of the space (pragma insertion, code transformation, and data caching) are inherently interconnected. Inserting unrolled pragmas influences array partitioning and thus the utilization of BRAM. The schedule and pragmas influence the possibility of caching data on-chip, and the choice of schedule impacts how we want to unroll based on loop properties (e.g., reduction loop), and so on. 
% Thus, the choice of one element affects the others, necessitating backtracking if the spaces are separated. This complexity complicates the exploration of the space and also requires constant verification of the legality of transformations.
% A reduction loop (e.g., loop L2 in Listing 
% % \ref{lst:gemm_motiv_ori}
% \ref{lst:original}
% ) is a loop in which a reduction operation is applied to an array. For the remainder of the paper, we will refer to loops that are not reduction loops as non-reduction loops (e.g., L0, L1, and L3 in Listing 
% % \ref{lst:gemm_motiv_ori}
% \ref{lst:original}
% ).

The various components of the exploration space—pragma insertion, code transformations, and data caching—are inherently interdependent. For example, unrolling pragmas affect array partitioning, which in turn influences BRAM utilization. Similarly, loop scheduling and pragmas impact the potential for on-chip data caching, while the choice of schedule dictates how loops should be unrolled based on their characteristics, such as whether they are reduction loops. A reduction loop is a loop where an operation, like addition or multiplication, accumulates values across iterations to produce a single output, often by reducing a multi-element array to a scalar (e.g., loop L2 in Listing \ref{lst:original}). The choice of schedule for reduction loops, as opposed to non-reduction loops (e.g., L0, L1, and L3 in Listing \ref{lst:original}), influences unrolling strategies and overall optimization because of the dependencies inherent to the loop’s characteristics.

% Consequently, selecting one element impacts the others, often requiring backtracking when the spaces are treated separately. This interconnection adds complexity to the exploration process and necessitates ongoing verification to ensure the legality of transformations.

Selecting one optimization affects others, often requiring backtracking when spaces are handled separately. This interdependence complicates exploration and demands constant verification of transformation legality.

% \vspace{-0.2cm}

\begin{figure}[H]

% \vspace{-0.2cm}
        % \lstset{basicstyle=\fontsize{7}{10}\selectfont\ttfamily}
        \begin{lstlisting}[label={lst:original},caption={Original gemm Code}]
for (i = 0; i < 200; i++) { // L0
  for (j = 0; j < 220; j++) // L1
    C[i][j] *= beta; // S0
  for (k = 0; k < 240; k++) // L2
    for (j = 0; j < 220; j++)// L3
      C[i][j]+=alpha*A[i][k]*B[k][j]; } // S1\end{lstlisting}
% \vspace{-0.2cm}
        \begin{lstlisting}[label={lst:distributed},caption={Fully Distributed Original gemm Code}]
for (i = 0; i < 200; i++) // L0
  for (j = 0; j < 220; j++) // L1
    C[i][j] *= beta; // S0
for (i = 0; i < 200; i++) // L2
  for (k = 0; k < 240; k++) // L3
    for (j = 0; j < 220; j++) // L4
      C[i][j]+=alpha*A[i][k]*B[k][j]; // S1\end{lstlisting}
% \vspace{-0.2cm}
        \begin{lstlisting}[label={lst:strip},caption={Fully Distributed Original gemm Code with Loop Strip-Mined Twice}]
for (i0 = 0; i0 < I0_S0; i0++) // L0
 for (i1 = 0; i1 < I1_S0; i1++) // L1
  for (i2 = 0; i2 < I2_S0; i2++) // L2
    for (j0 = 0; j0 < J0_S0; j0++) // L3
     for (j1 = 0; j1 < J1_S0; j1++) // L4
      for (j2 = 0; j2 < J2_S0; j2++) // L5
        C[i][j] *= beta; // S0
for (i = 0; i < I0_S1; i++) // L6
 for (i = 0; i < I1_S1; i++) // L7
  for (i = 0; i < I2_S1; i++) // L8
    for (k = 0; k < K0_S1; k++) // L9
     for (k = 0; k < K1_S1; k++) // L10
      for (k = 0; k < K2_S1; k++) // L11
        for (j = 0; j < J0_S1; j++) // L12
         for (j = 0; j < J1_S1; j++) // L13
          for (j = 0; j < J2_S1; j++) // L14
            C[i][j]+=alpha*A[i][k]*B[k][j];// S1\end{lstlisting}
% \vspace{-0.5cm}
\end{figure}
\begin{figure*}[ht!]

        % \lstset{basicstyle=\fontsize{7}{10}\selectfont\ttfamily}
        \begin{minipage}{0.45\textwidth}
        \begin{lstlisting}[label={lst:small_mem},caption={Result found for gemm kernel using the NLP with the constraints: DSP = 2,000, 
On-chip mem = 320 kB}]
load_C(C, C_off_chip);
/**** Level 0 of S0 ****/
for (j0 = 0; j0 < 1; j0++) // L0
  for (i0 = 0; i0 < 4; i0++) // L1
/**** Level 1 of S0 ****/
    for (j1 = 0; j1 < 22; j1++) // L2
#pragma HLS pipeline
/**** Level 2 of S0 ****/
      for (j2=0; j2<10; j2++) { // L3
#pragma HLS unroll
        for (i2=0; i2<50; i2++) // L4
#pragma HLS unroll
          i = ...;
          C[i][j] *=beta; } // S0
/**** Level 0 of S1 ****/
for (k0 = 0; k0 < 48; k0++) // L5
  load_B_S1(B, B_off_chip, k0);
  load_A_S1(A, A_off_chip, k0);
  for (j0 = 0; j0 < 1; j0++) // L6
    for (i0 = 0; i0 < 4; i0++) // L7
/**** Level 1 of S1 ****/
      for (j1=0; j1 < 220; j1++) // L8
#pragma HLS pipeline
/**** Level 2 of S1 ****/
        for (j2=0; j2<1; j2++) // L9
#pragma HLS unroll
          for (i2=0; i2<50; i2++) // L10
#pragma HLS unroll
            for(k2=0;k2<5;k2++){ // L11
#pragma HLS unroll
              i = ...;
              C[i][j]+=alpha*A[i][k2]*B[k2][j];}// S1
store_C(C, C_off_chip);
\end{lstlisting}

\end{minipage}
\hspace{1cm}
\begin{minipage}{0.45\textwidth}

        % \lstset{basicstyle=\fontsize{7}{10}\selectfont\ttfamily}
        \begin{lstlisting}[label={lst:big_mem},caption={Result found for gemm kernel using the NLP with the constraints: DSP = 6,840
On-chip mem = 7.2 MB}]
load_C(C, C_off_chip);
load_A(A, a_off_chip);
load_B(B, B_off_chip);
/**** Level 0 of S0 ****/
for (i0 = 0; i0 < 1; i0++) // L0
  for (j0 = 0; j0 < 1; j0++) // L1
/**** Level 1 of S0 ****/
    for (j1 = 0; j1 < 55; j1++) // L2
#pragma HLS pipeline
/**** Level 2 of S0 ****/
      for (i2 = 0; i2 < 200; i2++) // L3
#pragma HLS unroll
        for (j2=0; j2<4; j2++) { // L4
#pragma HLS unroll
          i = ...;
          C[i][j] *= beta;} // S0
/**** Level 0 of S1 ****/
for (i0 = 0; i0 < 1; i0++) // L5
  for (j0 = 0; j0 < 1; j0++) // L6
    for (k0 = 0; k0 < 60; k0++) // L7
/**** Level 1 of S1 ****/
      for (j1 = 0; j1 < 220; j1++) // L8
#pragma HLS pipeline
/**** Level 2 of S1 ****/
        for (i2=0;i2<200;i2++) // L9
#pragma HLS unroll
          for (j2=0; j2<1; j2++) // L10
#pragma HLS unroll
            for(k2=0;k2<4;k2++){ // L11
#pragma HLS unroll
              i = ...;
              C[i][j]+=alpha*A[i][k]*B[k][j];}//S1
store_C(C, C_off_chip);
\end{lstlisting}
% % \vspace{-1cm}
\end{minipage}

% \vspace{-0.6cm}
\end{figure*}

% \vspace{-0.4cm}

\subsection{Limitation of the Current DSE Methods}

% Several frameworks \cite{heterocl, scalehls, zhaopolsca,hida,pouchet:fpga13,pom,chen2024allo} enable code transformations and pragma insertion, but their spaces are limited for both \cite{scalehls,zhaopolsca,hida,pouchet:fpga13,pom} or do not have automatic DSE to explore the space \cite{heterocl,chen2024allo}.
% % 
% Transformations are restricted to loop property-based permutations \cite{heterocl, scalehls, hida,pouchet:fpga13,pom} and/or the use of Pluto \cite{pluto}. Pluto is particularly efficient for CPU but does not account for FPGA-specific optimizations like pipelining or array partitioning and above all try to minimize the number of data transfer (Comm) which may limit parallelization and reduce performance on FPGA and are a different objective than us which try to minimize the global latency (Lat).
% % 
% On the other hand, numerous Design Space Explorations (DSEs) \cite{autodse, harp, sohrabizadeh2022gnn, nlp_dse} explore pragma insertion for a fixed loop order. Although the user may initially set the schedule, it can be difficult, even for experts, to predict which transformations are necessary prior to pragma insertion. 
% Furthermore, the multitude of code transformations renders the exploration of each pragma impractical in terms of scalability.
% % 
% The key differences of the frameworks are highlighted in Table \ref{tab:rw}.
% \input{sources/table_rw}

Several frameworks \cite{heterocl, scalehls, zhaopolsca,hida,pouchet:fpga13,pom,chen2024allo} enable code transformations and pragma insertion, but have limited exploration spaces \cite{scalehls,zhaopolsca,hida,pouchet:fpga13,pom} or lack automatic DSE \cite{heterocl,chen2024allo}. Transformations are often restricted to loop property-based permutations \cite{heterocl, scalehls, hida,pouchet:fpga13,pom} or rely on Pluto \cite{pluto}, which is optimized for CPUs but overlooks FPGA-specific techniques like pipelining and array partitioning. Additionally, Pluto minimizes memory transfer from off-chip to on-chip communication (Comm), which can limit parallelism on FPGAs, unlike our focus on reducing global latency (Lat).
Meanwhile, other DSE frameworks \cite{autodse, harp, sohrabizadeh2022gnn, nlp_dse} explore pragmas insertion for fixed loop orders, but predicting friendly transformations beforehand is challenging. Exploring every possible transformation is also impractical for large design spaces. The key differences between the frameworks are summarized in Table \ref{tab:rw}.

% % \vspace{-0.25cm}
% \begin{table}[H]
\begin{table}[!htb]
\footnotesize
    % \vspace{-0.2cm}
\centering
\begin{tabular}{@{}p{0.28\linewidth} p{0.05\linewidth} p{0.05\linewidth} p{0.05\linewidth} p{0.08\linewidth} p{0.05\linewidth}  p{0.05\linewidth} p{0.05\linewidth} p{0.05\linewidth} @{}}
\toprule
& Poly-Opt-HLS & POL-SCA/  Pluto             & Auto-DSE/ HARP & Hetero-CL / Allo & NLP-DSE & Scale-HLS/ POM & Sisy-phus\\
\midrule
Tiling               &  \yes    &  \yes & \no & \yes & \no & Limit. & \yes \\
Permutation             &  Limit.      &  Limit. & \no & \yes & \no & Limit. & \yes \\
% Only heuristic permutation &   \yes & \nogood & - &  \nogood & - & \yesbad &  \nogood \\
Pragma Insert. &  \yes &  \no & \yes & \yes & \yes & \yes & \yes \\
Code Trans + Pragma Insert. (unified) &  \yes   & \no & \no & \no & \no & \no &  \yes \\
% Unified optimization problem  & \no & \no & \no & \no & \no &  \yes \\
% Automatic DSE                 & \yes & \yes & \no & \yes & \yes & \yes \\
Off-chip comm. gen. &  \yes  & \yes & \yes & \yes & \yes & \no & \yes \\
% Automatic burst size selection & \no & \yes & \yes & \yes & \no & \yes \\ 
Enum. (AI, heuristics,...)&  \nogood & \nogood & \yesbad & Manual & \nogood & \yesbad & \nogood \\
Objective  & Comm & Comm & Lat & - & Lat & Lat & Lat \\
\bottomrule 
\end{tabular}%
% }
\caption{Comparison of different frameworks}
\label{tab:rw}
 % \vspace{-0.9cm}
\end{table}

% % \vspace{-0.9cm}

We will now examine examples that highlight the limitations of current methods.
For this, we use two different HLS DSE methods, AutoDSE \cite{autodse} and NLP-DSE \cite{nlp_dse,nlp_dse_poster}. A comparison with HARP \cite{harp} and ScaleHLS \cite{scalehls} will be presented in Section \ref{sec:evaluation}.
AutoDSE treats the source-to-source compiler Merlin \cite{merlin} and the associated HLS tools as a black box, adjusting pragmas based on the identified bottlenecks from previous iterations. In contrast, NLP-DSE employs Nonlinear Programming DSE, utilizing a lower bound-based objective function to achieve high QoR within a short timeframe.
Both of this DSE use the AMD source-to-source compiler Merlin \cite{merlin}. 
The compiler employs different code transformations such as strip-mining via the \textit{TILE} pragma or for partial loop unrolling. 
% However, it generally avoids permutations, except when the two innermost loops are partially unrolled. In this case, the compiler strip-mines these loops and permutes the innermost resulting fully unrolled loops.
It typically avoids permutations, except when partially unrolling the two innermost loops, in which case the compiler strip-mines these loops and applies permutations to the resulting fully unrolled ones. 
A more detailed description of this compiler can be found in \cite{nlp_dse}.

In the following examples, we use the HLS compiler Vitis 2023.2 with the unsafe-math option disabled. The targeted FPGA is the AMD/Xilinx Alveo U200. We evaluate two scenarios: the gemm kernel from Polybench \cite{polybench-web}, as shown in Listing 
% \ref{lst:gemm_motiv_ori}
\ref{lst:original}, and a convolutional neural network (CNN) layer where the problem size and original loop
order of CNN are I,J=256 (Output and Input Channels), H,W=224 (Height, Weight) and P,Q=5 (Filter Height and Weight).

The gemm kernel’s original loop order, as illustrated in Listing 1, has a non-reduction loop at the center of the second statement (L3).
This arrangement facilitates AutoDSE and NLP-DSE in unrolling this loop and pipelining the reduction loop (L2). 
In this scenario, if the reduction loop is pipelined \textbf{and} partially unrolled by a factor of $uf$, performance will degrade. While partial unrolling allows for parallel execution of the multiplications, it increases the pipeline depth due to the loop’s dependencies, as the $uf$ additions must still be performed sequentially. Consequently, when the reduction loop is pipelined, the initiation interval (II) will significantly increase, becoming a multiple of both $uf$ and the latency of the reduction operation. This occurs because the next iteration of the loop can only begin after the $uf$ additions are completed and the output is fully written.
% 
% Due to their inability to further increase parallelism because of the loop order, the designs yielding the best QoR for these two methods achieve a throughput of 20 gigaFLOPs per second (GF/s). 
% 
Their inability to further increase parallelism due to loop order limits the best designs to a throughput of 20 GFLOPs per second (GF/s).
% 
% Listing \ref{lst:gemm_motiv_autodse} highlights the design found by AutoDSE. 
% The pragmas with $factor=1$ or pipeline off are not utilized in the final design. However, they illustrate the potential space explored by AutoDSE with various factors.
In such instances, a transformation becomes imperative to augment parallelization.

A single CNN layer with 6 loops results in an enormous design space that cannot be explored exhaustively. Loop permutations alone represent 6! = 720 possibilities, and tiling is required to manage array sizes exceeding on-chip memory capacity.
While frameworks like AutoDSE and NLP-DSE achieve satisfactory throughput—42.15 GF/s and 31.80 GF/s, respectively—they are unable to explore the full design space, which includes both code transformations and pragma insertion, within a reasonable timeframe. For example, running AutoDSE for all possible loop permutations would take around 500 days, and for NLP-DSE, it would take 106 days.
Other frameworks, which apply code transformations based solely on loop properties, such as prioritizing reduction loops in the outermost positions, explore a much more restricted design space, limiting their potential for further optimization.
This underscores the challenge of navigating such a vast design space and highlights gaps in current approaches.

\subsection{Overview of \framework}

% Maybe precise what limitations??

% To overcome these limitations, we introduce \framework, a solution tailored for affine programs, where loop bounds and conditionals are affine functions of surrounding loop iterators and parameters (commonly referred to as Static Control Parts in the Polyhedral Model) \cite{10.1007/s10766-006-0012-3, Feautrier1988}. 
% Our objective is to define a search space that allows for the selection of various parameters, such as pragma insertion, loop ordering and tiling, within a single optimization problem. This approach avoids the need for backtracking or handling disjoint optimization problems.
% Our framework efficiently and rapidly explores this large design space, overcoming the limitations of other frameworks that either restrict the search space too much or explore a broad space with a DSE that only partially covers it. Most importantly, it guarantees that only valid code transformations are considered.
% To accomplish this, we propose a two-step process: first, the creation of a space tailored to each kernel, followed by the exploration of that space.

We introduce \framework, tailored for affine programs with loop bounds and conditionals as affine functions \cite{10.1007/s10766-006-0012-3, Feautrier1988}. 
It defines a unified search space for parameters like pragma insertion, loop ordering, and tiling, avoiding backtracking or disjoint optimizations. \framework efficiently explores this design space, overcoming limitations of frameworks that either overly restrict or partially explore it. Crucially, it ensures only valid transformations. The process involves creating a kernel-specific space and exploring it.

The first step involves creating a template with fully distributed code that includes three levels of loops, each targeting specific optimizations. To achieve this, we strip-mined each original loop twice, ensuring that one of these loops is included at each level (assuming the permutation is valid). As a result, each original loop can be utilized across all three levels.
For instance, in Listing \ref{lst:original}, the loop $k$ (L2) is used 
in level 0 (L5) and in level 2 (L11) in Listing \ref{lst:small_mem}.
Specifically, we have an innermost level for fine-grained unrolling, a level dedicated to pipelining, and another level focused on on-chip data caching based on available resources. 
As demonstrated in Listings \ref{lst:small_mem} and \ref{lst:big_mem}, depending on the resource constraints specified by the user, we obtain two different code versions. 
For instance, a reduced on-chip memory requirement necessitates the partial transfer of arrays $A$ and $B$ (lines 17, 18 in Listing \ref{lst:small_mem}), leading to the use of a smaller on-chip buffer size that meets the user’s constraints.

Once this template is generated, it establishes a search space where the unknowns consist of the loop bounds, loop ordering, buffer sizes for on-chip arrays, and the locations in the code where data transfers take place.
Although other DSE methods can explore this space, 
we chose to use a cost model formulated as a Nonlinear Programming (NLP) problem as in NLP-DSE \cite{nlp_dse}. 
This approach allows us to efficiently navigate the theoretical space within seconds or minutes, leveraging accurate modeling made possible by compile-time analysis for affine programs. 
It is important to emphasize that exhaustive exploration using a simple cost model or synthesizing each design with HLS \cite{autodse} would be impractical due to the sheer size of the search space. In contrast, the NLP approach enables us to explore this space quickly and efficiently, allowing for a more comprehensive investigation than traditional methods.

Sisyphus outperforms NLP-DSE \cite{nlp_dse, nlp_dse_poster} by exploring a larger optimization space, including tiling and loop permutations, unlocking performance unattainable by NLP-DSE, which requires separate runs for each transformation. Automating the entire process, Sisyphus reduces exploration time and delivers superior results. It generates Vitis HLS pragmas with broader parameter support, like array partitioning, enhancing accuracy and design space coverage. Its high model precision enables the generation of at most one or two optimized designs, avoiding NLP-DSE’s extensive iterations.

% \textcolor{red}{Diff with NLP-DSE end}

For gemm, \framework significantly enhances parallelism by splitting and permuting loops without relying on the reduction loop. In this configuration, the NLP identifies the setup  shown in Listing \ref{lst:big_mem}. Both pipelined loops achieve an initiation interval (II) of 1, resulting in a throughput of 210 GFLOPs per seconds (GF/s).
The reduction loop (L11) is partially unrolled, allowing the multiplications to be executed in parallel while performing four additions sequentially. This small unrolling, selected by the NLP, deepens the pipeline, improving the quality of results (QoR) even though the reduction must remain sequential.
% 
% For the CNN, the NLP selects all parameters within the design space, including loop order after exploring 6! possibilities, data transfer points, on-chip array size, and the level of parallelism. This comprehensive analysis takes just 13 minutes and yields a throughput of 341 GF/s.
For the CNN, the NLP selects all parameters in the design space, including loop order (exploring 6! possibilities), on-chip array size, parallelism levels, etc. This analysis completes in 13 minutes, achieving a throughput of 341 GF/s.
% The design discovered by \framework is showcased in Listing \ref{lst:cnn}.

% Thanks to a well-defined space that only considers legal transformations and exploration guided by a cost model formulated as an NLP, our framework is able 
% to select the theoretical optimum in the entire space
% in a matter of seconds or minutes and generate a design.
Thanks to a well-defined space that considers only legal transformations and exploration guided by a cost model formulated as an NLP, our framework can identify the theoretical optimum across the entire space and generate a design within seconds or minutes.

% \input{sources/cnn}
% \vspace{-0.2cm}

\section{Unified Space}
\label{sec:code_transformation}

We proceed to develop a methodology that integrates code transformation, tile size selection and pragma insertion into a unified optimization framework. This involves implementing maximal distribution for code transformation, followed by the design and implementation of a template capable of executing code transformation, tiles size selection and pragma insertion simultaneously. We illustrate our code transformation process step by step using the example from Listings \ref{lst:original} to \ref{lst:big_mem}.

\myparagraph{Input}
% \framework takes an affine C/C++ file as input and automatically extracts a polyhedral representation of the code using PoCC \cite{pocc-web}. This representation allows us to gather all relevant information from the code, such as loop order, loop bounds, array sizes, and more. We will use this information to create the template and to automatically generate the NLP discussed in Section \ref{sec:nlp}. 
% Our framework will be particularly effective for codes that are fully distributable and fully permutable. However, in cases where the transformations described later are not legally possible, our framework still functions but restricts the solution space more.
% 
\framework takes an affine C/C++ file as input and extracts a polyhedral representation using PoCC \cite{pocc-web}, gathering information like loop order, bounds, and array sizes. This data is used to create the template and generate the NLP. 
% It works best for fully distributable and permutable codes but restricts the solution space if transformations are not fully legal.
% Our framework will be particularly effective for codes that are fully distributable and fully permutable. However, in cases where the transformations described later are not legally possible, our framework still functions but restricts the solution space more.
The framework is most effective for fully distributable and permutable codes but still works with restricted solution space when transformations are not fully legal.

\myparagraph{Maximal Distribution}
% To kickstart the process, we begin by thoroughly distributing the program, with the goal of maximizing the distance between each statement. This distribution strategy creates ample opportunities for parallelization by maximizing the distance between dependencies (reverse of fusion).
% % 
% Typically, this results in one statement per loop body with perfectly nested loops. To achieve this, we employ ISCC \cite{iscc}.
% It allows us to explore various schedule distributions and validate the legality of transformations by ensuring dependency constraints are preserved. 
% The code presented in Listing \ref{lst:distributed} displays the fully distributed version of the code from Listing \ref{lst:original}.
 % 
% 
To initiate the process, we distribute the program to maximize the distance between statements, promoting parallelization by reducing dependency overlap. Typically, this yields one statement per loop body within perfectly nested loops. Using ISCC \cite{iscc}, we explore scheduling options and verify transformation legality by preserving dependencies. Listing \ref{lst:distributed} shows the fully distributed version of the code from Listing \ref{lst:original}.

\myparagraph{Strip-mining}
After maximal distribution, we apply strip-mining to each loop twice. This process transforms the original loop, such as the loop L0 in Listing \ref{lst:distributed}, into three loops with trip counts $I0\_S0$, $I1\_S0$, and $I2\_S0$ (L0, L1, L2 in Listing \ref{lst:strip}). The product of these three loop trip counts equals the original trip count, i.e., $I0\_S0 \times I1\_S0 \times I2\_S0 = 200$.
Since strip mining is inherently legal, there is no necessity to validate the legality of this transformation. 
Subsequently, we can arrange the loops in different permutations if it is legal.
After applying three levels of strip-mining, the code from Listing \ref{lst:distributed} is transformed into the code seen in Listing \ref{lst:strip}.

% \input{sources/gemm_template}

% \input{sources/gemm_strip}

% Following the explanation provided in the next paragraph, strip mining allows us to create opportunities for fine-grained unrolling, pipelining, sequential execution,  coarse-grained unrolling and tiling if loop permutation is legal. This expands the range of possibilities, allowing each loop to integrate diverse hardware directives and be reorganized across different levels of the code.

\myparagraph{Creation of different levels}
Next, we consider possible permutations of the strip-mined loops. If these permutations are legal, we establish three loop levels. 
We ensure legality by checking dependency preservation using ISCC \cite{iscc}.
\myparagraphee{Level 2:} The innermost level facilitates fine-grained unrolling (complete unrolling), which increases parallelism by duplicating statements and can utilize a tree reduction if a reduction loop is unrolled and the option is enabled. For example, this is illustrated by loops L9, L10, and L11 in Listing \ref{lst:big_mem}.
\myparagraphee{Level 1:} 
The middle level facilitates pipelining, enhancing throughput by overlapping loop iterations. 
In cases where loop permutations allow us to achieve perfectly nested loops suitable for pipelining, we restrict pipelining to a single loop to ensure efficient synthesis.
Flattening the loops at Level 1 into a single loop and then pipelining the resulting loop introduces significant design complexity and leads to excessively long synthesis times.
Thus, if a loop at Level 1 is pipelined, it implies that the trip counts of the other loops at the same level are reduced to one.
% , ensuring manageable design complexity and synthesis times. 
For instance, in Listing \ref{lst:big_mem}, the loop $j$ (L8) is pipelined for the statement S1. This results in the loops $k$ and $i$ (L10 and L11 in Listing \ref{lst:template} ) having a trip count of 1, enabling them to be eliminated from the code.
\myparagraphee{Level 0:} 
Finally, the outermost level manages coarse-grained unrolling, tiling, and/or sequential execution. It not only controls parallelism at a coarse level but also determines the on-chip buffer size through tiling.
As described in the following paragraph, the loop order at this level is included in the design space.
This level is illustrated by loops L5, L6, and L7 in Listing \ref{lst:small_mem}. The loop bounds determine the on-chip buffer sizes for arrays A and B, transferred below loop L5. Resource constraints also influence the sequential execution of loop L7.
If the loop body contains a single statement and loop permutation is feasible, we disable coarse-grained unrolling. This achieves equivalent results since our template allows fine-grained unrolling of the innermost loop.

\myparagraph{Loop Permutation}
Loop order selection is not required for levels 1 and 2, as only a single loop is pipelined, and the innermost level (level 2) is fully unrolled, making the loop order irrelevant.
Conversely, at the outermost level (level 0), \textbf{all permutations are considered}, and the NLP selects the permutation. 
In Listings \ref{lst:small_mem} and \ref{lst:big_mem}, we can see that the NLP selected two different permutations for the statement S1 (loops L5, L6, and L7).

% % \vspace{-0.6cm}

% % \vspace{-0.2cm}

\myparagraph{Fusion}
% \textbf{Fusion}
We have opted not to incorporate fusion into our model for several reasons. 
Firstly, our objective is to maximize parallelization for computation-bound kernels. 
Fusion may reduce or minimize the dependency distances between 
statements, 
limiting 
potential 
parallelization opportunities. 
Additionally, since our focus is on computation-bound kernels, the primary bottleneck lies in computation. Therefore, we are willing to incur a minor cost, even if it involves transferring a tiled array multiple times.

\myparagraph{Overal Template}
In the context of our example, these transformations yield the code presented in Listing 
\ref{lst:template}, with each level corresponding to specific transformations.
The iterator of the loop retains its original name, with an added number indicating the level. For example, loop L14 in Listing \ref{lst:template} refers to loop L2 in Listing \ref{lst:original} at level 2 of the strip mining.
Notably, our framework is capable of preserving the original loop order, allowing it to retain the existing loop structure if the code is already optimized.

\myparagraph{Design Space}
% Therefore, the challenge involves determining the trip counts of each loop (e.g., $J0\_S0$) and the loop order of the level 0.
% Additionally, the design must adhere to resource constraints, including the usage of DSP and on-chip memory. Hence, the trip counts corresponding to fully unrolled loops must be constrained to avoid over-utilization of DSP resources. The on-chip buffers size must not exceed the on-chip memory capacity. Consequently, for a large problem size, the outermost level will enable control over the buffer sizes.
The challenge is to determine loop trip counts (e.g., $J0\_S0$) and level 0 loop order while meeting DSP and on-chip memory constraints. 
Hence, the trip counts corresponding to fully unrolled loops must be constrained to avoid over-utilization of DSP resources. The on-chip buffers size must not exceed the on-chip memory capacity. Consequently, for a large problem size, the outermost level will enable control over the buffer sizes.

\begin{figure}[H]
% \vspace{-0.5cm}
    \begin{minipage}{0.50\textwidth}
        % \lstset{basicstyle=\fontsize{7}{10}\selectfont\ttfamily}
        \begin{lstlisting}[label={lst:template},caption={Template automatically generated for the gemm input code (Listing \ref{lst:original})}]
int perm_S0[2] // Permutation order (0 for i0, 1 for j0)
int perm_S1[3] //Array to store permutation order (i0,j0,k0)
int lb_S0[2] = {I0_S0, J0_S0}; // loop bounds for S0
int lb_S1[3] = {I0_S1, J0_S1, K0_S1}; // loop bounds for S1
// possibilities to cache on-chip A, B and C
/********* Level 0 of S0 *********/
for (L0_S0=0; L0_S0 < lb_S0[perm_S0[0]]; L0_S0++) // L0
// possibilities to cache on-chip C
 for (L1_S0=0; L1_S0 < lb_S0[perm_S0[1]]; L1_S0++) // L1
 // possibilities to cache on-chip C
/********* Level 1 of S0 *********/
  for (j1=0; j1 < J1_S0; j1++) // L2
   for (i1=0; i1 < I1_S0; i1++) // L3
#pragma HLS pipeline
/********* Level 2 of S0 *********/
    for (i2=0; i2 < I2_S0; i2++) // L4
#pragma HLS unroll
     for (j2=0; j2 < J2_S0; j2++) // L5
#pragma HLS unroll
      C[i][j]*=beta; // S0
/********* Level 0 of S1 *********/
for (L0_S1=0; L0_S1 < lb_S1[perm_S1[0]]; L0_S1++) // L6
// possibilities to cache on-chip A, B and C
 for (L1_S1=0; L1_S1 < lb_S1[perm_S1[1]]; L1_S1++) // L7
 // possibilities to cache on-chip A, B and C
  for (L2_S1=0; L2_S1 < lb_S1[perm_S1[2]]; L2_S1++) // L8
  // possibilities to cache on-chip A, B and C
/********* Level 1 of S1 *********/
   for (j1=0; j1 < J1_S1; j1++) // L9
    for (k1=0; k1 < K1_S1; k1++) // L10
     for (i1=0; i1 < I1_S1; i1++) // L11
#pragma HLS pipeline
/********* Level 2 of S1 *********/
      for (i2=0; i2 < I2_S1; i2++) // L12
#pragma HLS unroll
       for (j2=0; j2 < J2_S1; j2++) // L13
#pragma HLS unroll
        for (k2=0; k2 < K2_S1; k2++) // L14
#pragma HLS unroll
         i = i0 * I1_S1 * I2_S1 + i1 * I2_S1 + i2;
         j = j0 * J1_S1 * J2_S1 + j1 * J2_S1 + j2; 
         k = k0 * K1_S1 * K2_S1 + k1 * K2_S1 + k2; 
         C[i][j]+=alpha*A[i][k]*B[k][j]; // S1\end{lstlisting}
    \end{minipage}
% \vspace{-0.55cm}
\end{figure}

\section{NLP}
\label{sec:nlp}

% Presented here is a comprehensive set of constraints and variables employed in a nonlinear program aimed at discovering the theoretical solution space outlined in Section \ref{sec:code_transformation}. We employ the methodology proposed by \cite{nlp_dse, nlp_dse_poster}, adapted to our specific context. Similar to their approach, we establish an estimation of the latency by assuming optimistic DSP utilization, assuming perfect resource reuse. However, we afford the user the flexibility to adjust the DSP limit or opt for a pessimistic DSP utilization scenario, where no reuse between statements is considered. 
% Moreover, users have the ability to adjust the size of on-chip memory, define the maximum number of array partitions, and customize the latency and resources allocated for each operation. This flexibility ensures adaptability across various platforms and compilers.

% We present constraints and variables for a nonlinear program to explore the theoretical solution space in Section \ref{sec:code_transformation}, based on \cite{nlp_dse, nlp_dse_poster}. 
We automatically generate constraints and variables for a nonlinear program to explore the theoretical solution space outlined in Section \ref{sec:code_transformation}, adapting the approach from \cite{nlp_dse, nlp_dse_poster} to our context.
Latency is estimated with optimistic DSP utilization, assuming perfect resource reuse, but users can adjust the DSP limit or choose a pessimistic scenario with no reuse. Users can also customize on-chip memory size, maximal array partitioning, and latency/resources per operation, ensuring adaptability to different platforms and compilers.
To gather all the required information for the NLP, we utilize PoCC \cite{pocc-web} to extract the intermediate representation.

\subsection{Variables}

% Table \ref{tab:var} 
Table \ref{tab:constant}
defines the sets, variables, and constants utilized in our NLP formulation. 

% % \vspace{-0.4cm}

% \input{sources/table_var_split}

% % \vspace{-0.6cm}

% \input{sources/table_var_split}

\subsection{Constraints}

% We now describe the precise meaning and implications of each constraint. To make this more understandable, we provide examples using the listings presented in the paper.
We describe each constraint’s meaning and implications with examples from the paper’s listings.

\myparagraph{Trip Count}
Equation \ref{eq:TC} constrains the trip count of each loop, ensuring that the product of the trip counts equals the original trip count.
The three loops after strip-mining, e.g., L6, L7, and L8 in Listing \ref{lst:strip}, must be equivalent to the original loop L2 in our example from Listing \ref{lst:distributed}. Therefore, we require $I0\_S0 \times I1\_S0 \times I2\_S0 = 200$.
% \vspace{-0.4cm}

\begin{eqnarray}
\label{eq:TC}
 \forall l \in \mathcal{L}, \prod_{v \in \mathcal{V}} TC_{l, v} = TC_l
\end{eqnarray}

% % \vspace{-0.2cm}

% \begin{table}[!htb]
\begin{table}[H]
% \vspace{-0.2cm}
\centering
% \resizebox{\textwidth}{!}{%
\begin{tabular}{@{}p{0.18\linewidth}p{0.80\linewidth}@{}}
\toprule
Set & Description \\
\midrule
$\mathcal{L},\mathcal{A},\mathcal{S}$ & the set of loops, arrays and statements \\
$\mathcal{S}_{1}$ & the set of statements alone in a loop body \\
$\mathcal{O}_s$ & the list of operations of the statements $s$ \\
% $\mathcal{O}_{s_{par}}$ & the operation which can be done in parallel, i.e., does not have any loop-carried dependence \\ 
$\mathcal{L}_s$ &  the set of nested loops which iterate the statement $s$ \\
$\mathcal{L}_s^{red}$ &  the set of reduction loops iterating the statement $s$ \\
$\mathcal{C}_{a_d}$ & the set of loops iterating the array $a$ at dimension $d$ \\ 
$\mathcal{V}$ & 
Strip mining level: 0 for coarse-grained/sequential, 1 for pipeline, 2 for fine-grained unrolled \\
% Level of the strip mining, 0 for coarse-grained/sequential, 1 for pipeline and 2 for fine-grained unrolled \\
$AP_{a,d}$ & Array Partition for the array $a$ in dimension $d$ \\
\midrule
Constant & Description \\
\midrule
$TC_l$ & Trip Count of the loop $l$ before strip-mining \\
$II_l$ & II of the loop $l$ \\
$IL_{par}, IL_{red}$ & Iteration Latency of the operations without ($par$) and with ($red$) dependencies of the statement $s$\\
% $IL_{red}$ & Iteration Latency of the operations with dependencies of the statement $s$\\
$DSP_{s_{op}}$ & Number of DSP used for the statement $s$ for the operation $op$ (This number accounts for the number of times the statement is replicated due to unrolling)\\
$DSP_{available}$ & Number of DSP available for the FPGA used \\
$max_{part}$ & Maximum array partitioning defined by the user \\
$ft\_arr_a\_loop_l$ & Footprint of the array $a$ if transferred to on-chip after the loop $l$ \\
$reuse_{opt}$ & Boolean for optimistic reuse \\
% \bottomrule
% \end{tabular}%
% }
% \caption{Overview of the set, constant and variables employed in formulating the NLP}
% \label{tab:var}
% \end{table}
\midrule
% \begin{table}[!htb]
% \centering
% % \resizebox{\textwidth}{!}{%
% \begin{tabular}{@{}p{0.16\linewidth}p{0.84\linewidth}@{}}
% \toprule
% \midrule
% \toprule
Variable & Description \\
\midrule

$tc_{l, level}$ & TC of the loop $l$ for the level of strip-mining \\
$loop_l\_UF$ & Coarse-grained unroll factor of the loop $l$ at level 0 \\
% of the strip-mining\\
$loop_l\_pip$  & Boolean to know if the loop $l$ is pipelined at level 1 \\
% of the strip-mining\\
$cache_l\_arr_a$ & 
Boolean to know if the array $a$ is transferred on-chip after the loop $l$ at level 0 
\\
% of the strip-mining \\
$perm_s$ & ID of the permutation of the statement $s$ \\
$burst_a$ & The maximum bitwidth at which the array $a$ can be transferred from off-chip to on-chip \\
\bottomrule
\end{tabular}%
% }
\caption{Overview of the sets, constants and variables employed in formulating the NLP}
\label{tab:constant}
% \vspace{-0.65cm}
\end{table}

% % \vspace{-1.2cm}

\myparagraph{Pipeline}
Constraints 
% \ref{eq:pip}, \ref{eq:pip2}, \ref{eq:one_pip_per_nesteed_loop}, and \ref{eq:II} 
\ref{eq:pip}--\ref{eq:II}
facilitate the selection of loop pipelining.
Constraint \ref{eq:pip} ensures each loop has a flag indicating if it is pipelined. 
For example, in Listings \ref{lst:small_mem} and \ref{lst:big_mem}, $S1$ shows loop $j1$ (L8) is 
pipelined, while its original loop is $j$ (L3 in Listing \ref{lst:original}),
making $loop_{j}\_pip=1$. 
As per Section \ref{sec:code_transformation}, only one loop at level 1 can be pipelined, requiring the sum of pipeline flags must be less than or equal to 1 (Eq. \ref{eq:one_pip_per_nesteed_loop}).
Non-pipelined loops must have a trip count of 1 (Eq. \ref{eq:pip2}). For instance, in Listing \ref{lst:template}, only loop L9 is pipelined at level 1 of $S1$, so $loop_j\_pip = 1$, while
$loop_i\_pip = 0$ and $loop_k\_pip = 0$. Loops L10 and L11 have trip counts of 1 and can be removed.

Eq. \ref{eq:II} calculates the initiation interval (II) based on loop properties and reduction operation latency. In Listings \ref{lst:small_mem} and \ref{lst:big_mem}, pipelined loops are non-reduction loops with no dependencies, giving $II = 1$. However, if loop L10 in Listing \ref{lst:template} were pipelined, it would involve a reduction requiring $II \geq IL_+$, as the write of iteration $k$ must complete before the read of $k+1$.

\begin{eqnarray}
\label{eq:pip}
\forall l \in \mathcal{L}, loop_l\_pip \in \{0,1\}
\\
\label{eq:pip2}
\forall l \in \mathcal{L}, (1-loop_l\_pip) \times TC_{l,1} == 1
% \end{eqnarray}
% \begin{eqnarray}
\\
\label{eq:one_pip_per_nesteed_loop}
% \begin{cases}
\forall s \in \mathcal{S}, \sum_{l \in \mathcal{L}_s} loop_l\_pip \leq 1
\\
\label{eq:II}
\forall s \in \mathcal{S}, II_s = \sum_{l \in \mathcal{L}_s} loop_l\_pip \times II_l 
\end{eqnarray}

% \vspace{-0.1cm}

\myparagraph{Coarse-grained unrolling}
% We enforce coarse-grained unrolling solely for non-reduction loops (Eq. \ref{eq:uf_if_seq}). For example, the loop L3 in Listing \ref{lst:distributed} cannot be coarse-grained unrolled as it is a reduction loop.
% % 
% Additionally, the unroll factor (UF) must divide the trip count of the current loop and be less than or equal to it (Eq. \ref{eq:uf_divide_tc} and \ref{eq:loop_uf_less_than_tc}). 
% However, as elaborated in Section \ref{sec:code_transformation}, coarse-grained unrolling is disregarded if there is only one statement within the loop (Eq. \ref{eq:uf_if_perfectly}). This implies that for Listing \ref{lst:original} 
% coarse-grained unrolling is disabled.
% 
Coarse-grained unrolling applies only to non-reduction loops (Eq. \ref{eq:uf_if_seq}). For instance, loop L3 in Listing \ref{lst:distributed} cannot be unrolled as it is a reduction loop.
The unroll factor (UF) must divide the loop trip count and not exceed it (Eq. \ref{eq:uf_divide_tc} and \ref{eq:loop_uf_less_than_tc}). However, as noted in Section \ref{sec:code_transformation}, coarse-grained unrolling is skipped if the loop has only one statement (Eq. \ref{eq:uf_if_perfectly}), as in Listing \ref{lst:original}.

\begin{eqnarray}
% % \vspace{-0.5cm}
\label{eq:uf_if_seq}
\forall s \in \mathcal{S}_1, \forall l \in \mathcal{L}_s^{red}, loop_l\_UF = 1
\\
\label{eq:uf_divide_tc}
% \begin{cases}
\forall l \in \mathcal{L}, loop_l\_UF \% TC_{l,0} == 0 \\
\label{eq:loop_uf_less_than_tc}
 \forall l \in \mathcal{L}, loop_l\_UF \leq TC_{l, 0} 
\\
\label{eq:uf_if_perfectly}
\forall s \in \mathcal{S}_1, \forall l \in \mathcal{L}_s, loop_l\_UF = 1 
\end{eqnarray}
% \vspace{-0.4cm}

\myparagraph{On-chip memory}
% Constraints \ref{eq:cache} and \ref{eq:only_one} ensure that each array, represented by a boolean (Eq. \ref{eq:cache}), is cached on-chip at a single location within the same loop body, as only one boolean can be true for each loop body (Eq. \ref{eq:only_one}). 
% % 
% In Listing \ref{lst:template}, for instance, array $A$ can be cached either before any loops (line 5) or after L6, L7, or L8 (lines 23, 25, and 27). In Listing \ref{lst:big_mem}, $A$ is cached on-chip with the on-chip footprint matching the size of the original array (line 2). Meanwhile, in Listing \ref{lst:small_mem}, $A$ is cached on-chip at line 18, but with a smaller on-chip footprint compared to the original.
% % 
% Based on where each array is transferred, we can compute the on-chip footprint. For example, in Listing \ref{lst:small_mem}, array $A$ is transferred below loop L5, which has a trip count of 48, so the footprint of $A$ on-chip is $200 \times \frac{240}{48}$.
% 
Constraints \ref{eq:cache} and \ref{eq:only_one} ensure each array, represented by a boolean (Eq. \ref{eq:cache}), is cached at a single location per loop body, with only one boolean true per loop (Eq. \ref{eq:only_one}).
In Listing \ref{lst:template}, array $A$ can be cached before loops (line 5) or after L6, L7, or L8 (lines 23, 25, 27). In Listing \ref{lst:big_mem}, $A$ is cached on-chip, matching the original array’s size (line 2). In Listing \ref{lst:small_mem}, $A$ is cached on-chip at line 18 with a smaller footprint.
The on-chip footprint depends on the transfer location. For Listing \ref{lst:small_mem}, $A$ is transferred below loop L5 (trip count 48), so its footprint is $200 \times \frac{240}{48}$.

% \vspace{-0.5cm}

\begin{eqnarray}
\label{eq:cache}
\forall l \in \mathcal{L}, \forall a \in \mathcal{A}, cache_l\_arr_a \in \{0,1\}
\\
\label{eq:only_one}
\forall a \in \mathcal{A}, \forall s \in \mathcal{S}, \sum_{l \in \mathcal{L}_s}, cache_l\_arr_a = 1
\\
\label{eq:foot}
 \sum_{a \in \mathcal{A}} \sum_{l \in \mathcal{L}} cache_l\_arr_a   \times ft\_array_a\_loop_l \leq Mem
\end{eqnarray}

% \vspace{-0.2cm}

\myparagraph{Array partitioning}
% We consider unrolling only if all the data needed for unrolling can be accessed in parallel, i.e., these data are on different on-chip banks thanks to the pragma \textit{array\_partitioning}. For each array $a$ and each dimension $d$ of this array, we want the factor of the cyclic array partitioning for this dimension to be greater than the unroll factor (Eq. \ref{eq:max_part3}) to ensure that all data are on different banks.
% % 
% If the array is reused, we need to ensure that the array partitioning is optimal for each statement. Therefore, we force the array partitioning to be a multiple of each unroll factor that iterates through this dimension (Eq. \ref{eq:max_part2}). $AP_{a,d}$ gives the cyclic array partitioning factor for array $a$ in dimension $d$ that we use in the generated code. Eq. \ref{eq:max_part1} assures that the array partitioning per array (product of the partitioning factors per dimension) is less than or equal to the limit fixed by the user.
% % 
% For example, in Listing \ref{lst:big_mem}, $C$ is unrolled 200 times in the first dimension (lines 11, 12) and 4 times in the second dimension (lines 13, 14) for the first loop body, and 200 times in the first dimension (lines 25, 26) for the second loop body. Thus, we have $AP_{C,1} = 200$ and $AP_{C,2} = 4$, and $200 \times 4 = 800 \leq max_{part}$, where $max_{part}$ is 1,024 in our case.
% 
Unrolling is applied only if all required data can be accessed in parallel, ensured by pragma \textit{array\_partitioning}. For each array $a$ and dimension $d$, the cyclic partitioning factor must exceed the unroll factor (Eq. \ref{eq:max_part3}) to ensure data is on separate banks.
If the array is reused, partitioning must align with each unroll factor iterating through the dimension (Eq. \ref{eq:max_part2}). $AP_{a,d}$ represents the cyclic partitioning factor used in the code, with the total partitioning per array constrained by the user-defined limit (Eq. \ref{eq:max_part1}).
For example, in Listing \ref{lst:big_mem}, $C$ is unrolled 200 times in the first dimension (lines 11, 12) and 4 times in the second (lines 13, 14) for the first loop body, and 200 times in the first dimension (lines 25, 26) for the second. This gives $AP_{C,1} = 200$, $AP_{C,2} = 4$, and $200 \times 4 = 800 \leq max_{part}$, with $max_{part} = 1,024$.

% \vspace{-0.5cm}

\begin{eqnarray}
\label{eq:max_part3}
\forall a \in \mathcal{A}, \forall d \in \mathbb{N}, \forall l \in C_{a_d}, 
AP_{a,d} \geq TC_{l,2} \\
\label{eq:max_part2}
\forall a \in \mathcal{A}, \forall d \in \mathbb{N}, \forall l \in C_{a_d}, 
AP_{a,d} \% TC_{l,2} == 0  \\
\label{eq:max_part1}
\forall a \in \mathcal{A}, \prod_{d \in \mathbb{N}} 
AP_{a,d} \leq max_{part} 
\end{eqnarray}

% \vspace{-0.1cm}

\myparagraph{DSP utilization}
% Finally, we need to manage DSP utilization. Given the challenges in predicting DSP reuse, we use a boolean flag, $reuse_{opt}$, set by the user to determine the level of DSP reuse. If $reuse_{opt} = 1$, we assume an optimistic reuse scenario: DSPs used in one loop body can be directly reused by subsequent loop bodies if the operations are identical. However, if the operations differ, as discussed in \cite{nlp_dse, nlp_dse_poster}, no reuse is assumed. Therefore, the total DSP usage is calculated as the sum of the maximum DSPs required for each operation across all loop bodies (Eq. \ref{eq:dsp_optimistic}). If $reuse_{opt} = 0$, we assume no reuse between different loop bodies, and we sum the DSP requirements for each statement (Eq. \ref{eq:dsp_pessimist}).
% Based on the value of $reuse_{opt}$, the total DSP usage must be less than or equal to the available DSPs (Eq. \ref{eq:dsp1} and \ref{eq:dsp2}).
To manage DSP utilization, we use a user-defined flag, $reuse_{opt}$. If $reuse_{opt} = 1$, optimistic reuse is assumed: identical operations reuse DSPs across loop bodies, with total usage given by the maximum DSPs per operation (Eq. \ref{eq:dsp_optimistic}). However, if the operations differ, as discussed in \cite{nlp_dse, nlp_dse_poster}, no reuse is assumed. If $reuse_{opt} = 0$, no reuse is assumed, and DSPs are summed for each statement (Eq. \ref{eq:dsp_pessimist}). In both cases, total usage must not exceed available DSPs (Eq. \ref{eq:dsp1} and \ref{eq:dsp2}).
% 
% For instance, if a multiplication operation ($*$) uses 3 DSPs and an addition operation ($+$) uses 2 DSPs, in Listing \ref{lst:big_mem}, S0 requires $200 \times 4 \times 3$ DSPs for multiplications, and S1 requires $200 \times 4 \times 3 + 200 \times 4 \times 3 + 200 \times 4 \times 2$ DSPs for two multiplications and one addition.
% % 
% With optimistic reuse for the same operation (e.g., multiplication), the DSP usage is calculated as: $\max(200 \times 4 \times 3, 2 \times 200 \times 4 \times 3) + 200 \times 4 \times 2$.
% % 
% With pessimistic reuse, the DSP usage is: $200 \times 4 \times 3 + 2 \times (200 \times 4 \times 3) + 200 \times 4 \times 2$.
If a multiplication ($*$) uses 3 DSPs and an addition ($+$) uses 2 DSPs, in Listing \ref{lst:big_mem}, S0 needs $200 \times 4 \times 3$ DSPs for multiplications, while S1 requires $200 \times 4 \times 3 + 200 \times 4 \times 3 + 200 \times 4 \times 2$ DSPs for two multiplications and one addition.
With optimistic reuse, DSP usage for multiplication is $\max(200 \times 4 \times 3, 2 \times 200 \times 4 \times 3) + 200 \times 4 \times 2$, while pessimistic reuse results in $200 \times 4 \times 3 + 2 \times (200 \times 4 \times 3) + 200 \times 4 \times 2$.

% \vspace{-0.5cm}
\begin{eqnarray}
\label{eq:dsp_optimistic}
 DSPs\_used_{opt} = 
     \sum_{op \in \{+,-,*,/ \}} \max_{s \in \mathcal{S}}(DSP_{s_{op}} / II_s) 
\\
\label{eq:dsp_pessimist}
 DSPs\_used_{pes} = 
     \sum_{op \in \{+,-,*,/ \}} \sum_{s \in \mathcal{S}}(DSP_{s_{op}} / II_s) 
\\
\label{eq:dsp1}
reuse_{opt} \times DSPs\_used_{opt}  \leq DSP_{available} \\
\label{eq:dsp2}
 (1-reuse_{opt}) \times DSPs\_used_{pes} \leq DSP_{available}
\end{eqnarray}

% \vspace{-0.2cm}

\subsection{Objective Function}

We use the objective function similar to the approach described in \cite{nlp_dse, nlp_dse_poster}, tailoring it to meet the specific needs of our problem. Here, $Lat^s_2$ corresponds to the fine-grained unrolled level (level 2) for the statement $s$, $Lat^s_1$ denotes the pipeline tile that incorporates the fine-grained unrolled level (level 1), and recursively, $Lat_0$ encompasses $Lat_1$, enabling coarse-grained unrolling and facilitating on-chip data caching (level 0).

The memory latency $Lat^s_{mem}$ for a statement $s$ corresponds to the time required to transfer the array from off-chip to on-chip and back for the arrays needed by a statement $s$. We assume the memory transfer is pipelined, allowing the transfer of one data element (with a maximum bitwidth of 512 bits) per cycle. Therefore, the latency at a given level is calculated as the array footprint divided by the burst size (bitwidth used for transferring the array), multiplied by the trip count of the loop that iterates over the function responsible for the transfer, based on the selected permutation of level 0.
Additionally, if an array is fully transferred on-chip for a statement $s$ and reused by a subsequent statement $s{’}$, no further transfers are needed.
The burst size for an array must divide its last dimension. For instance, if we have an array $float \text{ } A[512][20]$, the maximum burst size for this array is $128$ bits. This is because $128$ is the largest power of $2$ that can evenly divide the size of the last dimension ($20 \times 32$ bits).

% \vspace{-0.4cm}

\begin{equation*}
\left.
\begin{cases}
& Lat^s_2 = IL_{par} +
IL_{seq} \times \prod_{l\in \mathcal{L}^{red}}  \times \log_2(TC_{l,2})  \\

 & Lat^s_1 = 
Lat^s_2 + II \times (TC_{l,1} - 1) \\

    & Lat^s_0 = \prod_{l\in \mathcal{L}} \frac{TC_{l,0}}{loop_l\_UF} \times Lat^s_1
 \\

& L^s_{mem} = \sum_{l \in \mathcal{L}}   \max_{a\in \mathcal{A}} (cache_{l}\_arr_a \times ft\_array_a\_loop_l \\
& \;\;\;\;\;\;\;\;\;\;\;\;\;\;\;\;\;\;\;\;\;\;\;\;\;\;\;\;\; \;\;\;\;\; \times TCs(perm_s) / burst_a)
\\
& Lat_s = Lat^s_0 + L^s_{mem}
    \end{cases}
    \right.
\end{equation*}
% Finally, the objective function 
% % (Eq. \ref{eq:obj}) 
% is defined as the sum of the latencies of each loop body $obj\_func = \sum_{s \in \mathcal{S}} Lat_s$.

% \begin{equation}
% \label{eq:obj}
% obj\_func = \sum_{s \in \mathcal{S}} Lat_s
% \end{equation}

% \vspace{-0.4cm}

Finally, the objective function 
% \begin{equation}
% \label{eq:obj}
$obj\_func = \sum_{s \in \mathcal{S}} Lat_s$
% \end{equation}
is defined as the sum of the latencies of each loop body.

% % \vspace{-2cm}

\section{Post-Optimization and Customization}
\label{sec:code_generation}

% We generate the code using the results obtained from the NLP solver, which provide us with the trip count for each loop and level of strip-mining, as well as information about which loops are pipelined, the array partitioning for each array and dimension and where the array are cache. Additionally, we incorporate further optimizations to enhance the Quality of Results (QoR).

% \subsection{Code Generation}

% The NLP allows us to find the trip count of each loop, the array partitioning for each array, and the size of the on-chip buffers. By considering the loop order and the pragmas applied to each loop as described in Section \ref{sec:code_transformation}, we can directly generate the code. We automatically generate the functions that load data from off-chip to on-chip with the maximum burst size possible and vice versa with the functions that store the data off-chip. Additionally, we incorporate further optimizations to enhance the QoR.

% The NLP determines the trip count for each loop, array partitioning, and on-chip buffer sizes. 
% Functions for loading/storing data between off-chip and on-chip memory are automatically generated with the maximum burst size. Further optimizations are also applied to improve QoR.
The NLP determines the trip counts, array partitioning, and on-chip buffer sizes. Data transfer functions between off-chip and on-chip memory are automatically generated with the maximum burst size. Further optimizations are also applied to improve QoR.

\myparagraph{Optimization for Non-Constant Trip Count}
% \label{sec:optimi_nn_cte}
% 
% 
To optimize loops with non-constant trip counts, we implement a code transformation that preserves the NLP's estimation while simplifying compilation for the HLS compiler. We achieve this by replacing loops with non-constant trip counts with the maximal trip count computed using PoCC \cite{pocc-web}. 
% Subsequently, we replace all the statement iterate by this loop by the call to a function e.g, instead of $C[i][i] += B[i][i]$ we replace by $C[i][j] = f(C[i][j], B[i][i])$ where the function ensure compliance with the constraints of the non-constant trip count loop. This function returns the computation if the constraints are met or returns just the value of the output otherwise.
Next, we replace all statements iterated by this loop with calls to a function. For example, instead of using \( C[i][i] += B[i][i] \), we replace it with \( C[i][j] = f(C[i][j], B[i][i]) \), where the function $f$ ensures compliance with the constraints of the non-constant trip count loop. This function returns the computed value if the constraints are satisfied; otherwise, it simply returns the original output value.
This technique helps to reduce compilation time and achieve designs with a good QoR.

% Introducing a condition solely above the statement results in excessive compilation times. Therefore, we employ these techniques to reduce compilation time and achieve designs with a good QoR.

\myparagraph{Grouping Data Transfers}
% \label{sec:overlaping}
% 
In the NLP, we already account for parallel transfers of arrays when they are transferred at the same level in the code (e.g., within the same loop). However, we do not consider parallel transfers between arrays belonging to different loop bodies. 
In our example, we first manage the transfer of arrays required for initializing the matrix, transferring the arrays for multiplication only after the initialization is complete. However, FPGA DRAM banks allow parallel transfers of different arrays. Consequently, we can optimize load and store operations during post-processing to improve memory transfer overlap.

\myparagraph{Code Transformation for HLS}
% \label{sec:code_trans_hls}
% 
% To simplify the understanding of the HLS compiler, particularly to ensure that the compiler can pipeline loops with the correct initiation interval (II), we simplify the reductions (regardless of whether we use tree reductions) in the fine-grained unrolled part (level 2). Therefore, under the pipeline we create a variable to accumulate the reduction, which we then add to the output.
To help the HLS compiler pipeline loops with the correct initiation interval (II), we simplify reductions at level 2. A variable accumulates the reduction within the pipeline, which is then added to the output.
Additionally, we use the \textit{loop\_flatten off} pragma on all reduction loops above the pipeline. Although this adjustment could theoretically improve the QoR, in practice, flattening these loops can complicate the code and potentially increase the initiation interval (II) of the pipeline.

\myparagraph{Customization}
Our framework is designed to be explainable and user-friendly, requiring the user to provide only the affine C/C++ code of the kernel to be optimized and the available resources.
To fully benefit developers, our NLP formulation allows the template to be completely customized to the user’s needs. Each variable in the design space can be defined by the user if they have specific constraints.
For example, in the gemm template (Listing \ref{lst:template}), the user can set the permutations for level 0 while allowing the NLP to explore other parameters. Additionally, they can specify the on-chip size of matrix $A$ by directly defining a variable in the NLP, allowing for exploration within the defined constraints.

% \vspace{-0.2cm}

\section{Evaluation}
% \label{sec:evaluation}

% The primary goal of this evaluation is to demonstrate how well our approach can apply the appropriate code transformations and insert the hardware directives while respecting the resource constraints define by the user.
% To do this, we compared our method with four other frameworks: AutoDSE \cite{autodse}, NLP-DSE \cite{nlp_dse}, HARP \cite{harp}, and ScaleHLS \cite{scalehls}. 

% As all the framework does not allow to generate bitstream without important human intervention, we first demonstrate our ability to find a design with good QoR by using all the resources of the board and generating HLS report with \textit{vitis-flow} which allow to consider memory transfer from off-chip to on-chip which oblige us to add memory transfer to ScaleHLS code.

% Hence we show that our designs are synthetizable. We evaluate only on a subset of the kernel due to the time to generate bitstream.

The goal of this evaluation is to demonstrate how effectively our approach applies code transformations and inserts hardware directives while respecting user-defined resource constraints.
We compared our method against four other frameworks: AutoDSE \cite{autodse}, NLP-DSE \cite{nlp_dse}, HARP \cite{harp}, and ScaleHLS \cite{scalehls}.
Since generating a bitstream without significant manual intervention is challenging for some frameworks, we first demonstrate our ability to achieve a high-quality design by utilizing all available resources—ensuring fair comparison with the other frameworks—without considering place and route. We generate HLS reports using the \textit{vitis-flow}, which accounts for memory transfers between off-chip and on-chip, necessitating the inclusion of memory transfer steps in ScaleHLS’s code.
Thus, we show that our designs are synthesizable by targeting the resources of a single SLR, which eliminates the need for manual partitioning. Due to the time required for bitstream generation, we restrict our evaluation to a subset of the kernels.

% The goal of this evaluation is to demonstrate how effectively our approach applies code transformations and inserts hardware directives while respecting user-defined resource constraints. We compared our method to four other frameworks: AutoDSE \cite{autodse}, NLP-DSE \cite{nlp_dse}, HARP \cite{harp}, and ScaleHLS \cite{scalehls}. Since bitstream generation without manual intervention can be challenging for some framework, we first demonstrate our ability to achieve high-quality designs by fully utilizing available resources. We generate an HLS report using the \textit{vitis-flow}, which includes memory transfers, adding these steps to ScaleHLS’s code. We show our designs are synthesizable by targeting a single SLR, avoiding manual partitioning. Due to bitstream generation time, we limit the evaluation to a subset of kernels.

%  Explain about SLR why 1 slr

\label{sec:evaluation}

% % \vspace{-0.2cm}

\subsection{Setup}
\label{sec:eval:setup}

% We conducted our experiments using kernels from Polybench/C 4.2.1 \cite{polybench-web}, along with a Convolutional Neural Network (CNN) kernel and matrix multiplication tasks similar to those used in the BERT transformer model. The computations were performed with single-precision floating-point data types to allow for a direct comparison with frameworks like AutoDSE, NLP-DSE, and ScaleHLS. The datasets used are medium-sized ones from Polybench/C \cite{polybench-web}.
We conducted experiments using kernels from Polybench/C 4.2.1 \cite{polybench-web}, a CNN kernel, and matrix multiplication tasks similar to those in the BERT transformer model. All computations used single-precision floating-point data for direct comparison with frameworks like AutoDSE, NLP-DSE, and ScaleHLS, using medium-sized Polybench/C datasets.
% 
% In addition, we compared our approach with HARP \cite{harp}, which integrates a Graph Neural Network (GNN) model to predict the behavior of HLS tools. HARP explores the design space for one hour using this model and synthesizes the top-10 designs. For our evaluation, we used medium-sized Polybench kernels with double-precision floating-point data to simulate the best conditions, as detailed in \cite{nlp_dse}. It’s important to note that HARP requires specific fine-tuning to perform effectively with various problem sizes.
% 
% 
% We also compare our approach with HARP \cite{harp}, which seeks to enhance High-Level Synthesis design exploration by integrating a Graph Neural Network model that predicts HLS tool behavior. They conduct exploration within this space utilizing the GNN model for one hour and then synthesizing the top-10 designs. 
% We use the medium-sized Polybench kernels found in their training set with double-precision floating-point data types, aiming to deploy HARP under optimal conditions. As detailed in \cite{nlp_dse}, HARP's effectiveness is not universally applicable without the necessary fine-tuning or training, particularly when confronted with diverse problem sizes.
We also compare our approach with HARP \cite{harp}, which enhances HLS design exploration using a Graph Neural Network (GNN) model to predict HLS tool behavior. HARP explores this space for one hour and synthesizes the top 10 designs. We use the medium-sized Polybench kernels found in their training set with double-precision floating-point data types, aiming to deploy HARP under optimal conditions. As detailed in \cite{nlp_dse}, HARP's effectiveness is not universally applicable without the necessary fine-tuning or training, particularly when confronted with diverse problem sizes.

We selected these problem sizes to highlight how our method performs in transforming code and inserting pragmas in cases where full unrolling is not feasible. We also demonstrate the effectiveness of our framework in tiling large tasks such as CNN and bert\_3072\_100, ensuring that only a small portion (a tile) of each array fits on-chip, in line with memory constraints.
For the CNN kernel, the problem size and loop dimensions 
% are I, J = 256 (output and input channels), H, W = 224 (height and width), and P, Q = 5 (filter dimensions). 
are the same as described in Section \ref{sec:motivation}.
For bert\_n\_m matrix multiplication, the dimensions are I = n, J = m, K = n (reduction).

% We used the AMD/Xilinx Vitis HLS compiler \cite{vitis} to demonstrate the effectiveness of our method. Report generation was performed using Vitis 2023.2. When enabling tree reduction, we used the “funsafe math optimizations” option, which allows commutative and associative reduction operators to enable reductions in logarithmic time. 
% For HARP comparisons, we followed the settings outlined by the authors, using Vitis 2021.1 and enabling tree reduction.
We used the AMD/Xilinx Vitis HLS compiler \cite{vitis} to demonstrate our method’s effectiveness, generating reports with Vitis 2023.2. When enabling tree reduction, we selected the “funsafe math optimizations” option to allow commutative and associative reduction operators for logarithmic time reductions.
% 
% For our hardware setup, we targeted the Xilinx Alveo U200 device running at 250 MHz. 
% We use the commercial solver Gurobi 11.0.0 \cite{gurobi} with AMPL. The experiments were performed on an AMD EPYC 7V13 64-core processor, with 32 GB of memory and 4 cores allocated for each run.
For our hardware setup, we targeted the Xilinx Alveo U200 device operating at 250 MHz. We utilized the commercial solver Gurobi 11.0.0 \cite{gurobi} with AMPL. 
% The experiments were conducted on an AMD EPYC 7V13 64-core processor, equipped with 32 GB of memory, with 4 cores allocated per run.

% \vspace{-0.1cm}

\subsection{Experimental Evaluation}

To ensure a fair comparison, we modified kernels with non-constant trip counts for all benchmarks, as detailed in Section \ref{sec:code_generation}, except for ScaleHLS, which already handles such loops.
We generated AutoDSE’s design space using the \textit{ds\_generator} command, applying all trip count factors for unroll and tile sizes. Since AutoDSE lacks array partitioning, its design space is slightly larger. AutoDSE was run with a 1,000-minute exploration timeout and a 180-minute HLS synthesis timeout. For NLP-DSE, we followed the original paper’s parameters \cite{nlp_dse}. HARP was run with the authors’ settings \cite{harp} using Vitis 2021.1 and tree reduction enabled.
ScaleHLS assumes that all kernel data are already located in on-chip memory. To provide a fair comparison, we modified their code to handle off-chip to on-chip memory transfers. We optimized these transfers to 512 bits per cycle, with all arrays transferred in parallel. Each array was assigned to a different DRAM bank. \textit{However, due to the extensive manual modifications required, we did not extend this comparison to all kernels.}
% 
% 
% For \framework evaluation, we set the Gurobi solver timeout to 14,400 seconds (4 hours) per design. By default, we assume optimist resource reuse between non-parallel statements. If DSP utilization is exceeded, we rerun the NLP with pessimistic assumptions (no resource reuse) and regenerate the design. The maximum partitioning constraint, $max_{part}$, was set to 1024, following AMD/Xilinx’s compiler and FPGA guidelines.
For \framework evaluation, we set the Gurobi solver timeout to 14,400 seconds per design. By default, we assume optimistic resource reuse between non-parallel statements. If DSP utilization exceeds limits, we rerun the NLP with pessimistic assumptions (no resource reuse) and regenerate the design. The maximum partitioning constraint, $max_{part}$, was set to 1024, per AMD/Xilinx guidelines.

Each synthesis starts with a C-simulation to verify correctness. We generate Vitis reports using the \textit{vitis-flow}, which counts memory transfers for all frameworks and considers the resources of the entire FPGA. For bitstream generation, we consider only one SLR to eliminate any manual intervention. We consider 60\% of the resources of a single SLR to quickly identify a synthesizable design, minimizing the number of times AutoDSE needs to be run.

% \vspace{-0.1cm}

\subsection{Comparison of the Methods}

\label{sec:comparison}

\subsubsection{AutoDSE, NLP-DSE and Scale-HLS with tree reduction}

\label{sec:eval_hls}

Table \ref{tab:eval_tree} provides a comparison among AutoDSE (A-DSE), NLP-DSE (N-DSE), Scale-HLS (S-HLS) and \framework (ours) with tree reduction. 
% 
% The \textbf{TR} column denotes the status of the tree reduction option, while 
% 
The \textbf{T} column indicates the specific transformations applied, with D representing distribution, T for data tiling, and P for permutation.
The column \textbf{Perf. Impr. of ours framework vs.} displays the performance improvement of our framework compared to the three different frameworks.
% 

% \input{sources/tablebis}

% % \vspace{-0.3cm}
% \begin{table}[H]
\begin{table}[H]
% \begin{table}[!htb]
% \vspace{-0.3cm}
% \footnotesize % TODO idk if i have right
% \begin{adjustwidth}{-1in}{-1in}
    
\centering
\begin{tabular}{@{}l c  r  rr r  r @{}}
\toprule \\[-3.0ex]
&  &  \multicolumn{1}{l}{GF/s} & & \multicolumn{3}{l}{Perf. Impr. of ours framework vs.}  \\[-0.4ex]
\cmidrule{3-3}
\cmidrule{5-7}  \\[-3.0ex]

\textbf{Kernel} &  \textbf{T} & 
% \textbf{AutoDSE} & \textbf{NLP-DSE} & 
\textbf{Ours} & & \textbf{A-DSE} & \textbf{N-DSE} & \textbf{S-HLS}   \\[-0.4ex]

\midrule
2mm & D,P & 175.49 &  & 431.17x & 1.42x & 5.50x\\
3mm & D,P & 140.83 &  & 80.75x & 1.02x & 3.86x \\
atax & D & 1.96 &  & 0.99x & 1.00x & 1.23x \\
bicg & D,P & 1.96 &  & 1.97x & 1.97x & 1.18x\\
doitgen & D,P & 54.05 &  & 1.35x & 2.71x & 8.45x\\
gemm & D,P & 203.39 &  & 1.84x & 1.55x & 5.81x\\
gemver & P & 17.62 &  & 4.58x & 1.74x & 9.39x\\
gesummv & D & 1.96 &  & 0.99x & 1.00x & 0.71x \\
mvt & P & 13.24 &   & 1.70x & 1.70x & 1.26x\\
symm & D,P & 240.50 &  & 8.28x & 8.28x & 2,704.51x\\
syr2k & D,P & 254.69 &  & 5.52x & 1.86x & 174.47x\\
syrk & D,P & 161.20 &  & 6.55x & 2.32x & 358.95x\\
trmm & D,P & 148.53 &  & 8,345.10x & 8,312.42x & 2,376.91x \\
\midrule \\[-3.0ex]
Average &   & 108.88 &  & 683.91x  & 641.46x & 434.78x \\
Geo Mean &  & 40.13 &  & 9.17x  & 3.47x & 16.15x \\

\bottomrule

\end{tabular}
\caption{Throughput (GF/s) Comparison 
% of \framework, AutoDSE, NLP-DSE, and ScaleHLS 
with Tree Reduction Using Vitis HLS}

\label{tab:eval_tree}
% \end{adjustwidth}
% \vspace{-0.9cm}
\end{table}

Across all evaluated kernels, we consistently achieve performance that is comparable to or better than alternatives, except for the atax, and gesummv kernels, where we observe a slowdown of less than 5\%.
For 3mm, symm and trmm we had to apply the constraint indicating no reuse (Eq. \ref{eq:dsp_pessimist}), as the constraint with optimistic reuse (Eq. \ref{eq:dsp_optimistic}) resulted in kernels with resource over-utilization.
For Gesummv, ScaleHLS is 1.41x faster. However, 91\% of our design’s execution time is spent on data transfers. ScaleHLS benefits from 512-bit data transfers, which we manually applied for testing. Our framework uses 64-bit chunks, the largest bitwidth that divides the array. With 512-bit transfers, our framework could achieve 14.50 GF/s.
For other memory-bound kernels like Atax and Gesummv, the slight performance differences observed with other frameworks are largely due to how Merlin handles memory transfers. 

% % \vspace{-0.05cm}

\framework averages a speedup of 683.91x, 641.46x and 434.78x over AutoDSE, NLP-DSE and Scale-HLS respectively, across the evaluated kernels. In terms of geometric mean, \framework achieves a speedup of 9.17x, 3.47x and 16.15x.

% \myparagraph{Code which does not need code transformation}

\myparagraph{Preserve the original schedule when necessary}
For atax, \framework preserves the original schedule and inserts pragmas, effectively maintaining the existing schedule while efficiently incorporating pragmas, similar to two DSEs focused on pragma insertion.

% \myparagraph{Code transformation to manage reduction loop}
% In the cases of 2mm, bert, and gemm, we observe significant improvements due to code transformations, especially in effectively managing the reduction loop.

% % This enhancement is particularly evident with bert\_3072\_100, where the reduction loop exhibits a large trip count, as well as in gemm when tree-reduction is not used.

% For the CNN layer, its large size leaves no choice but to transfer the \textit{input} array multiple times.

\myparagraph{Code transformation for memory-bound kernel}
For mvt and gemver, both memory-bound kernels, effective optimization entails loop permutations to prevent redundant array transfers and accommodate array partitioning within the constraints of the AMD/Xilinx compiler's 1024 limit. \framework adeptly manages these loop permutations and pragma insertions, guaranteeing optimized performance for these kernels. 
For bicg, fully distributing and arranging the loops appropriately can enhance parallelization capabilities.

\myparagraph{Allow to achieve a parallelism not achievable without loop transformation}
For doitgen, symm, syrk, syr2k, and trmm, the original schedule imposes constraints on the achievable level of parallelism. This limitation highlights the performance enhancements made possible by \framework. We observe QoR improvements for these kernels because the code transformations enable a combination of loop unrolling that cannot be achieved even with maximal loop distribution, as each loop can be partially unrolled.

\subsubsection{AutoDSE and NLP-DSE without tree reduction}

Transformations without tree reduction, which may be needed to preserve floating-point data precision, must be different since the optimizations differ when tree reductions are not involved. Table \ref{tab:eval_wtree} provides a comparison among AutoDSE (A-DSE), NLP-DSE (N-DSE) and \framework (ours) without tree reduction.

% % \vspace{-0.3cm}
% \begin{table}[H]
% \begin{table}[H]
\begin{table}[!htb]
% \vspace{-0.3cm}
% \footnotesize % TODO idk if i have right
% \begin{adjustwidth}{-1in}{-1in}
    
\centering
\begin{tabular}{@{}l c  r  r r  r @{}}
\toprule \\[-3.0ex]
&  &  \multicolumn{1}{l}{GF/s} & & \multicolumn{2}{l}{Perf. Impr. Ours vs.}  \\[-0.4ex]
\cmidrule{3-3}
\cmidrule{5-6}  \\[-3.0ex]

\textbf{Kernel} &   \textbf{T} & 
% \textbf{AutoDSE} & \textbf{NLP-DSE} & 
\textbf{Ours} & & \textbf{A-DSE} & \textbf{N-DSE}    \\[-0.4ex]

\midrule

2mm & D,P & 171.07 &  & 817.42x & 1.44x \\
3mm & D,P & 140.11 &  & 511.64x & 1.04x  \\
atax & D & 1.86 &  & 0.98x & 1.00x \\

bicg & D,P & 1.86 &  & 1.92x & 1.93x \\
doitgen & D,P & 54.05 &  & 5.46x & 2.71x \\
gemm & D,P & 210.61 &  & 9.25x & 4.01x \\
gemver & P & 15.01 &  & 37.28x & 8.73x \\

gesummv & D,P & 1.81 &  & 0.97x & 0.97x  \\

mvt & P & 9.75 &   & 1.42x & 1.42x \\

symm & D,P & 240.55 &  & 8.28x & 7.32x \\

syr2k & D,P & 429.34 &  & 17.61x & 3.14x \\

syrk & D,P & 333.03 &   & 13.54x & 4.78x \\

trmm & D,P & 130.90 &  & 7,362.68x & 7,333.85x  \\

\midrule \\[-3.0ex]
Average &   & 133.84 &  & 676.03x  & 567.10x  \\
Geo Mean &  & 41.62 &  & 18.50x  & 4.49x  \\

\bottomrule

\end{tabular}
\caption{Throughput (GF/s) Comparison 
% of \framework, AutoDSE and NLP-DSE 
without Tree Reduction Using Vitis HLS}

\label{tab:eval_wtree}
% \end{adjustwidth}
% \vspace{-0.9cm}
\end{table}

% 

% \framework achieves an average speedup of 676.03x over AutoDSE and 567.53x over NLP-DSE across the evaluated kernels without tree reduction. The geometric mean speedup is 18.50x and 4.80x, respectively.
% % 
% On average, the performance improvement without tree reduction is greater than with tree reduction, particularly for gemm (Listing \ref{lst:big_mem}) and gemver. This is because, without tree reduction, optimizing the reduction loop becomes challenging, making code transformations even more necessary.
% The difference between syrk and syr2k with tree reduction arises because the HLS compiler achieves an II=2 instead of the expected II=1. Without tree reduction, the compiler achieves II=1 as expected.

\framework achieves an average speedup of 676.03x over AutoDSE and 567.10x over NLP-DSE across kernels without tree reduction, with geometric means of 18.50x and 4.49x, respectively.
Performance improvement is higher without tree reduction, especially for gemm (Listing \ref{lst:big_mem}) and gemver, as optimizing reduction loops becomes harder, increasing the need for code transformations. The difference between syrk and syr2k with tree reduction occurs because the HLS compiler achieves II=2 instead of the expected II=1, while without tree reduction, II=1 is achieved.

\subsubsection{AutoDSE and NLP-DSE for Bert and CNN}

Table \ref{tab:eval_bert} compares AutoDSE (A-DSE), NLP-DSE (N-DSE), and \framework (ours).

The \textbf{TR} column shows the tree reduction status. 
For bert\_100\_64, we observed a minor decrease in QoR, attributed to the NLP selecting a configuration where the pipeline was not applied as expected.

% % \vspace{-0.3cm}
% \begin{table}[H]
\begin{table}[H]
% \begin{table}[!htb]
% \vspace{-0.3cm}
% \footnotesize % TODO idk if i have right
% \begin{adjustwidth}{-1in}{-1in}
    
\centering
\begin{tabular}{@{}l @{\hspace{-0.6mm}}rc  r @{\hspace{-0.6mm}} rr r   @{}}
\toprule \\[-3.0ex]
& & &  \multicolumn{1}{l}{GF/s} & & \multicolumn{2}{l}{Perf. Impr. Ours vs.}  \\[-0.4ex]
\cmidrule{4-4}
\cmidrule{6-7}  \\[-3.0ex]

\textbf{Kernel} &  \textbf{TR} & \textbf{T} & 
% \textbf{AutoDSE} & \textbf{NLP-DSE} & 
\textbf{Ours} & & \textbf{A-DSE} & \textbf{N-DSE}  \\[-0.4ex]
\midrule

bert\_100\_64 & 1 & D,P & 81.22 &  & 0.99x & 0.95x \\
bert\_100\_64 & 0 & D,P & 76.76 &  & 1.09x & 1.09x \\
bert\_100\_768 & 1 & D,P & 218.08 &  & 1.12x & 1.19x \\
bert\_100\_768 & 0 & D,P & 216.95 &  & 1.13x & 1.13x \\
bert\_100\_3072 & 1 & D,P & 241.59 &  & 1.21x & 1.21x \\
bert\_100\_3072 & 0 & D,P & 241.06 &  & 1.22x & 1.22x \\
bert\_3072\_100 & 1 & D,P,T & 314.89 &  & 21.85x & 24.29x \\
bert\_3072\_100 & 0 & D,P,T & 82.41 &  & 989.65x & 1,156.02x \\
cnn & 1 & D,P,T & 314.15 &   & 7.38x & 8.23x \\
cnn & 0 & D,P,T & 341.20 &  & 8.09x & 10.73x \\

\midrule \\[-3.0ex]
Average &  & & 212.83 &  & 103.37x  & 120.61x  \\
Geo Mean & & & 185.32 &  & 4.38x  & 4.69x  \\

\bottomrule

\end{tabular}
\caption{Throughput (GF/s) Comparison 
% of \framework, AutoDSE and NLP-DSE 
Using Vitis HLS}

\label{tab:eval_bert}
% \end{adjustwidth}
% \vspace{-0.9cm}
\end{table}

\myparagraph{Code transformation with tiling}
The selection of loop order and tile size is critical for determining which array sizes, particularly for large problem size like bert\_3072\_100 and CNN layers, should be cached on-chip and where they should be transferred. 
In these instances, the NLP aims to enhance data reuse and minimize unnecessary memory transfers while identifying the theoretical optimal balance with parallelism.

% % \vspace{-0.17cm}
\subsubsection{HARP}
In Table \ref{tab:harp} we compare the throughput and resource utilization (BRAM, DSP, LUT, FF) achieved with HARP \cite{harp}.

% % \vspace{-0.25cm}
% \begin{table}[H]
\begin{table}[!htb]
    % \vspace{-0.3cm}
\centering
\begin{tabular}{@{}l @{\hspace{-1.2mm}} r 
% @{\hspace{-.4mm}} 
r @{\hspace{-1mm}} l rr r @{}}
\toprule \\[-3.0ex]
&   \multicolumn{2}{l}{GF/s} & &  \multicolumn{2}{l}{Resource Utilization (\%)} & Perf. Impr. \\
&   & & &  \multicolumn{2}{l}{BRAM,DSP,LUT,FF} \\[-0.8ex]
\cmidrule{2-3}
\cmidrule{5-6}

\textbf{Kernel} &  \textbf{HARP} & 
% \textbf{AutoDSE} & \textbf{NLP-DSE} & 
\textbf{Ours} & & \textbf{HARP} & \textbf{\framework}  \\[-0.4ex]

\midrule \\[-3.0ex]
atax & 1.72 & 1.96 &&        78,52,49,50 & 38,65,53,33 & 1.14x \\
bicg & 0.92 & 1.96 &&        75,25,35,36 & 38,65,48,31 & 2.13x  \\
gemm & 125.59 & 87.86 &&     29,80,55,40 & 32,48,36,23 & 0.70x \\
gemver & 1.66 & 9.42 &&      29,28,35,19 & 34,8,31,11 & 5.69x \\
mvt & 7.07 & 8.29 &&         40,78,43,30 & 53,44,57,30 & 1.17x \\
\midrule \\[-3.0ex]
Average & 27.39 & 20.71 &  &  & & 1.45x\\
Geo Mean & 4.71 & 6.28 &  &  & & 1.33x\\
\bottomrule 
\end{tabular}%
% }
\caption{Comparison with HARP Using Vitis HLS}
\label{tab:harp}
 % \vspace{-0.8cm}
\end{table}

% % \vspace{-0.9cm}

The enhanced performance seen in both bicg and gemver can be attributed to the same underlying factor observed in the comparison with AutoDSE and NLP-DSE. Regarding mvt, HARP produced comparable performance to our own due to their use of the Merlin compiler, which facilitates loop permutations when partially unrolling consecutive loops. This allows them to find the same schedule that our framework identified. 
% For gemm, HARP’s design falls outside our parameter space due to restrictions on maximal array partitioning.
However, for gemm, the design discovered by HARP falls outside our specified parameter space, due to our restriction on maximum array partitioning.
% However, for gemm, the design discovered by HARP falls outside our specified parameter space, particularly because of our restriction on maximal array partitioning. 
% Notably, it is the only design in the top 10 that outperforms our framework.

\subsection{Single SLR Onboard Evaluation}

\label{sec:eval_on_board}

% \begin{table}[H]
\begin{table}[!htb]
% \vspace{-0.4cm}
\centering
\begin{tabular}{@{}l 
% @{\hspace{-1.2mm}} 
rr 
@{\hspace{-1mm}} 
r
r 
% rr 
r @{}}
\toprule \\[-3.0ex]
&
GF/s
&
% \multicolumn{2}{l}{GF/s} & 
&  \multicolumn{2}{l}{RU: BRAM,DSP,LUT(K),FF(K)} & Perf. Impr. \\
% &   
% % & &
% &  \multicolumn{2}{l}{BRAM,DSP,LUT(K),FF(K)} \\[-0.8ex]
% \cmidrule{2-3}
% \cmidrule{5-6}
\cmidrule{4-5}

\textbf{Kernel} &  \textbf{Ours} &
% \textbf{AutoDSE} & \textbf{Ours} & 
&\textbf{Ours} & \textbf{AutoDSE}    \\[-0.4ex]

% 1.03
% 1.03
% 44.32
% 29.89

\midrule \\[-3.0ex]
2mm & 30.57 & 
% 0.35 & 44.32 &
& 
510,556,213,276 & 
353.5,963,287,292 &  76.98x \\
3mm & 29.89 & 
% 0.39 & 29.89 &
&

611,984,230,300 
& 470,1117,278,306 & 72.34x  \\
atax & 1.03 & 
% 0.23 & 1.03 &
&     
450,173,240,250
& 630.5,452,170,212 & 4.63x \\
bicg & 1.02 & 
% 0.59 & 1.11 &
&
      
      451,173,238,265 
      & 867.5,196,168,217 &  1.80x \\
% \midrule \\[-3.6ex]
% Average & 27.39 & 23.83 &  &  & & 3.25x\\
% Geo Mean & 4.71 & 8.76 &  &  & & 1.86x\\
\bottomrule 
\end{tabular}%
% }
\caption{Comparison of Onboard Evaluation for 1 SLR}
\label{tab:bit}
% \vspace{-0.8cm}
\end{table}

  The resource utilization (RU) is detailed in the thousands for both LUTs and FFs. URAM utilization is excluded as no kernels use it. And the last column show the performance improvement we achieve with Sisyphus over AutoDSE. 
For the kernels evaluated on board, we can see that our generated design is both faster and consumes fewer resources compared to AutoDSE.

% % \vspace{-0.1cm}

% % \vspace{-0.2cm}

% \subsection{Latency estimation}

% Implementing the post-optimization strategy outlined in Section \ref{sec:overlaping} results in a loss of the lower bound discussed in \cite{nlp_dse_poster, nlp_dse}, primarily because we do not include communication overlap modeling between different loop bodies in the NLP. This omission is intentional, as incorporating such modeling would significantly increase the search time.

% Similarly, we opt against integrating the \textit{loop\_flatten} pragma into our model, despite its potential automatic invocation by the compiler for loops without reductions. As detailed in Section \ref{sec:code_generation}, we intentionally deactivate it for loops with reductions. Although this optimization's automatic application results in the loss of the lower bound, we have made this choice owing to its rare implementation. Our aim is to avoid reliance on an optimization that may not be consistently available.

% In Figure \ref{fig:accuracy}, we depict the estimated latency by the NLP of the evaluated designs  and juxtapose them with the latencies reported by HLS. 
% % 
% Our post-optimization analysis shows that, in most cases, the latencies estimated by NLP exceed those reported by HLS. 

% \begin{figure}[!htb]
%     \centering
%     \includegraphics[width=0.99\linewidth]{images/res.png}
%     \captionsetup{skip=1pt}
%     \caption{Comparison of the latency reported by HLS with the estimated latency obtained through NLP}
%     \label{fig:accuracy}
% \end{figure}

\subsection{Latency Estimation and Scalability}

% Implementing the post-optimization strategy from Section \ref{sec:code_generation} leads to a loss of the lower bound discussed in \cite{nlp_dse_poster, nlp_dse}, 
% primarily because we exclude communication overlap modeling between loop bodies in the NLP to prevent significantly increasing search time.
% We also choose not to incorporate the \textit{loop\_flatten} pragma into our model, even if the compiler may automatically invoke it for loops without reductions.
% As noted in Section \ref{sec:code_generation}, we deactivate it for reduction loops.
% While this results in a lower bound loss, we prefer not to rely on an optimization that may not be consistently implemented.

Implementing the post-optimization strategy from Section \ref{sec:code_generation} results in a loss of the lower bound discussed in \cite{nlp_dse_poster, nlp_dse}, mainly because we exclude communication overlap modeling between loop bodies in the NLP to avoid significantly increasing search time. We also do not incorporate the \textit{loop\_flatten} pragma, even if the compiler may automatically apply it for loops without reductions, as we deactivate it for reduction loops. Although this leads to a lower bound loss, we prefer not to depend on an optimization that may not be consistently implemented.
As noted in Section 5, our post-optimization focuses primarily on memory transfer overlap, so our final latency values do not directly reflect the model’s initial predictions. However, our model achieves (when predicting the latency of the final design generated) an average prediction error of 8\% across evaluated designs, with a 3\% error in geometric mean. For certain designs, like syrk and syr2k, where the compiler achieves an II=2 rather than the expected II=1, prediction errors are higher (49\% and 32\%, respectively) due to an optimistic estimation of II.

The average solving time of the NLP was 455 seconds, while the geometric mean stood at 6.7 seconds. Only the kernel 3mm with tree reduction timeout at 4h. 32/36 NLP's problem are solve in less than 1 minutes.
It is important to note that the solver provides a feasible solution early in the process, though it may not be theoretically optimal. As the search continues, the solution improves, and once the solver completes, the solution is theoretically optimal. Therefore, it is possible to set a timeout to obtain a feasible, but not optimal, solution more quickly.

% \vspace{-0.2cm}

\section{Related Work}
\label{sec:related}

% Design Space Exploration (DSE) methodologies for pragma insertion, such as those mentioned in \cite{nlp_dse_poster, nlp_dse, autodse, harp, sohrabizadeh2022gnn, lorenzo, comba, harp, lin, ironman,6560643,6850383,s2fa}, yield designs with a satisfactory QoR. However, they often require extensive computational time, with AutoDSE taking up to a day to complete.
% % 
% While users have the option to perform code transformations before conducting DSE, the process typically involves treating each code transformation and pragma insertion as separate optimization problems. This decomposition of the problem can result in a loss of QoR because predicting which transformation is necessary to achieve an optimal design is challenging.
% % 
% The NLP-DSE method, as explained in \cite{nlp_dse, nlp_dse_poster}, employs a non-linear solver to integrate pragmas into the code. We build upon the methodology developed by the authors and tailor it to our specific optimization space to determine the pragmas and schedule for the kernel.

Various code transformations have been devised for CPUs \cite{pluto},
% , pouchet.11.popl, kruse2020autotuning,baghdadi2019tiramisu}, 
GPUs \cite{ppcg},  FPGAs \cite{pouchet.fpga.13, zhaopolsca,zhao2021phism, li2014throughput, liu2016loop, liu2017polyhedral, 7160061, choi.iccad.18} or for all \cite{ansor}. While code transformations for CPUs and GPUs yield remarkable results tailored to their respective architectures, they may not be inherently suitable for FPGA targets aimed at achieving high parallelism. 
% Pluto \cite{pluto}, considered state-of-the-art, performs code transformations including tiling to minimize dependency distance between memory accesses, thereby facilitating data reuse. However, in our scenario, reducing dependency distance could potentially constrain parallelization efforts.
Regarding \cite{zhaopolsca,zhao2021phism}, they employ Pluto on different scopes of the kernel for code transformation, yet its pragma insertion capabilities are limited, making it incomparable to our work. Conversely, \cite{pouchet.fpga.13, li2014throughput, liu2016loop, liu2017polyhedral, 7160061, choi.iccad.18} pursue a distinct objective from ours. 
\cite{pouchet.fpga.13} aims to minimize communication between off-chip and on-chip, which results in better QoR than ours for memory-bound kernels. 
The works  \cite{li2014throughput, liu2016loop, liu2017polyhedral,7160061, choi.iccad.18} concentrate on code transformations aimed at enhancing pipelining techniques.
However, these objectives may not be suitable for computation-bound kernels requiring high levels of parallelization.

Ansor \cite{ansor} achieves impressive results for CPU and GPU architectures; however, there are notable differences with our approach. Ansor requires multiple evaluations to perform its design-space exploration (DSE) and update its cost model, which accelerates the DSE process but is not scalable for FPGA, where a single evaluation can take minutes or even hours. Furthermore, Ansor’s approach to code transformations and pragma insertions (which are more tailored for CPU/GPU) addresses these problems separately, potentially resulting in suboptimal designs.

% The \cite{scalehls, heterocl,pylog,chen2024allo,hida} compilers undertake code transformation and pragma insertion. However, these transformations are somewhat limited, mostly encompassing heuristic modifications based on loop properties. Moreover, the scope of pragma insertion is more limited compared to our proposed approach.

The choice of tile sizes significantly impacts the QoR. In line with our methodology, \cite{8056810, 10.1145/3445814.3446759} employ a cost model to determine the tile size. However, while \cite{8056810} focuses on minimizing communication overhead, our approach differs. On the other hand, \cite{10.1145/3445814.3446759} investigates tiles size selection for Convolutional Neural Networks (CNNs) with three-level CPU caching.

% % \vspace{-0.2cm}

\section{ Limitations}
\label{sec:limitation}

% As mentioned earlier, Sisyphus is designed to optimize affine loop nests. In cases where an application includes non-affine code segments, we can partition the code into affine and non-affine sections. This allows us to leverage the strengths of Sisyphus for affine regions, where it offers significant performance improvements. For non-affine sections, other DSE tools, such as AutoDSE \cite{autodse} or HARP \cite{harp}, can be utilized to achieve efficient optimizations. Once both affine and non-affine sections are optimized, we can integrate them using a dataflow model. This can be done either by connecting the sections with FIFOs for efficient data exchange or by employing TAPA \cite{tapa}, which facilitates dataflow programming. This combined approach ensures that both affine and non-affine portions of the application are handled in an optimal manner, achieving a balance between different types of optimizations and seamlessly integrating them for improved overall performance.

Sisyphus is designed to optimize affine loop nests. For applications with non-affine code, we can partition the code into affine and non-affine sections. Sisyphus handles affine regions effectively, while tools like AutoDSE \cite{autodse} or HARP \cite{harp} can optimize non-affine parts. The optimized sections are then integrated using a dataflow model, either with FIFOs for data exchange or by employing TAPA \cite{tapa}, which facilitates dataflow programming. This combined approach ensures that both affine and non-affine portions of the application are handled in an optimal manner, achieving a balance between different types of optimizations and seamlessly integrating them for improved overall performance.

% Our NLP formulation allows the user to fix a variety of parameters, such as loop order and buffer size. This enables the user to constrain the design space to meet specific requirements, like buffer size, while allowing Sisyphus to explore and optimize the code within these predefined constraints.

\section{Conclusion}
\label{sec:conclusion}

% In this paper, we introduced Sisyphus, a unified framework designed to streamline the process of hardware design through high-level synthesis. By integrating code transformation, pragma insertion, and tile size selection into a single optimization problem, our framework addresses the significant challenges posed by existing HLS workflows, which often rely on disjointed and manual methods.

% Our novel approach leverages Nonlinear Programming (NLP) to efficiently explore the vast design space of regular loop-based kernels, facilitating rapid identification of optimal transformations and pragma placements. The results of our evaluation demonstrate that Sisyphus not only achieves superior Quality of Results (QoR) compared to state-of-the-art frameworks such as AutoDSE, NLP-DSE, and ScaleHLS but also significantly reduces design generation complexity.

% In this paper, we introduced Sisyphus, a unified framework for streamlining hardware design through high-level synthesis. By combining code transformation, pragma insertion, and tile size selection into one optimization problem, Sisyphus addresses challenges in existing HLS workflows, which often rely on disjointed and manual methods.

% For this Sisyphus define a space sous forme de template.
% Using Nonlinear Programming (NLP), Sisyphus efficiently explores this space, quickly identifying transformations and pragma configuration to achieve a design with good QoR. Our evaluation shows that Sisyphus achieves superior QoR compared to AutoDSE, NLP-DSE, and ScaleHLS.

In this paper, we introduced Sisyphus, a unified framework that streamlines hardware design through HLS. By integrating code transformation, pragma insertion, and tile size selection into a single optimization, Sisyphus addresses challenges in existing HLS workflows, which often rely on manual and disjointed methods.

Sisyphus defines a design space in the form of a template. Using NLP, it efficiently explores this space, quickly identifying transformations and pragma configurations to achieve designs with good QoR. Our evaluation demonstrates that Sisyphus outperforms AutoDSE, NLP-DSE, and ScaleHLS in terms of quality of result.

\myparagraph{Acknowledgments - }
 This work was supported by the NSF award \#CCF-2211557. It is also supported
by CDSC industrial partners and
the AMD\footnote{J. Cong has a financial interest in AMD.}/HACC Program.

\clearpage
\bibliographystyle{ACM-Reference-Format}
% \bibliography{sample-base}
\balance
\bibliography{bibs/iccad23,bibs/sp,bibs/lnp,bibs/refs,bibs/gabriel,bibs/ierefs,bibs/ics15}

\end{document}